\documentclass[12pt,preprint]{aastex}

\def\kms{\ifmmode{\rm km\thinspace s^{-1}}\else km\thinspace s$^{-1}$\fi}

\usepackage{natbib}

\shorttitle{V\,12 in the open cluster NGC\,188}
\shortauthors{Meibom et~al.}

\begin{document}

\title{Age and distance for the old open cluster NGC\,188 from
the eclipsing binary member V\,12\altaffilmark{1}}

\author{S{\o}ren Meibom\altaffilmark{2,3}}
\affil{University of Wisconsin - Madison, WI, USA}

\author{Frank Grundahl}
\affil{Department of Physics and Astronomy, Aarhus University, Denmark}

\author{Jens Viggo Clausen}
\affil{Niels Bohr Institute, Copenhagen University, Denmark}

\author{Robert D. Mathieu}
\affil{University of Wisconsin - Madison, WI, USA}

\author{S{\o}ren Frandsen}
\affil{Department of Physics and Astronomy, Aarhus University, Denmark}

\and
\author{Andrzej Pigulski, Artur Narwid, Marek Steslicki}
\affil{Instytut Astronomiczny Uniwersytetu Wroclawskiego, Kopernika 11, 51-622 Wroclaw Poland }

\and
\author{Karolien Lefever}
\affil{Instituut voor Sterrenkunde, Katholieke Universiteit Leuven, 3001 Heverlee, Belgium}

\altaffiltext{1}{WIYN Open Cluster Study. XXXVII.}
\altaffiltext{2}{Presently at Harvard-Smithsonian Center for Astrophysics,
60 Garden Street, Cambridge, MA 02138, USA}
\altaffiltext{3}{smeibom@cfa.harvard.edu}


\begin{abstract}
We present time-series radial-velocity and photometric observations
of a solar-type double-lined eclipsing binary star (V\,12) in the
old open cluster NGC\,188. We use these data to determine the spectroscopic
orbit and the photometric elements for V\,12. From our analysis we
determine accurate masses ($M_{p} = 1.103 \pm 0.007~M_{\odot}$,
$M_{s} = 1.081 \pm 0.007~M_{\odot}$) and radii ($R_{p} = 1.424 \pm
0.019~R_{\odot}$, $R_{s} = 1.373 \pm 0.019~R_{\odot}$) for the primary (p)
and secondary (s) binary components. We adopt a reddening of $E_{B-V}=0.087$
for NGC\,188, and derive component effective temperatures of $5900
\pm 100\,K$ and $5875 \pm 100\,K$, respectively, for the primary and
secondary stars. From their absolute dimensions, the two components
of V\,12 yield identical distance moduli of $V_0-M_V = 11\fm24 \pm 0\fm09$,
corresponding to $1770 \pm 75$\,pc.
Both stars are near the end of their main-sequence evolutionary phase,
and are located at the cluster turnoff in the color-magnitude diagram.
We determine an age of $6.2 \pm 0.2$\,Gyr for V\,12 and NGC\,188,
from a comparison with theoretical isochrones in the mass-radius diagram.
This age is independent of distance, reddening, and color-temperature
transformations.
We use isochrones from Victoria-Regina and Yonsei-Yale with [Fe/H]
= -0.1 and [Fe/H] = 0.0. From the solar metallicity isochrones,
an age of 6.4\,Gyr provides the best fit to the binary components
for both sets of models. For the isochrones with [Fe/H] = -0.1,
ages of 6.0\,Gyr and 5.9\,Gyr provide the best fits for the Victoria-Regina
and Yonsei-Yale models, respectively. We use the distance and age
estimates for V\,12, together with best estimates for the metallicity
and reddening of NGC\,188, to investigate the locations of the
corresponding VRSS and $Y^2$ isochrones relative to cluster members in
the color-magnitude diagram. Plausible changes in model metallicity
and distance to better match the isochrones to the cluster sequences,
result in a range of ages for NGC\,188 that is more than 3 times
that resulting from our analysis of V\,12.
\end{abstract}



\section{Introduction}
\label{intro}

NGC\,188, once thought to be the oldest open cluster in our Galaxy
\citep[e.g.,][]{sandage62}, represents the old stellar population
of the Galactic disk. Due to its richness, location in the Galaxy
($\alpha_{2000} = 0^{h}~47^{m}$, $\delta_{2000} = +85\degr~15\arcmin$;
$l = 122.8\degr$, $b = 22.4\degr$), large population of single and
binary members \citep{pkm+03,gmh+08}, and its age, NGC\,188 is a
benchmark cluster for studies of stellar evolution and cluster dynamics,
and for the formation and chemical and dynamical evolution of the
Galactic disk. Determining an accurate and precise age for NGC\,188
is of general importance.

The age determined for NGC\,188 has changed dramatically since the early 
determination by \citet{sandage62} of 14-16 Gyr. Later studies derived
9-12 Gyr \citep{dl64,iben67}, and more recent values range from $\sim$6
Gyr \citep{ta89,ccc+90,dgg92,mmm93,ddg+95} to $\sim$7-8\,Gyr
\citep{hobbs90,shk+99,vs04}. Past age estimates were based on fitting
model isochrones to the cluster's main sequence, turnoff, and giant
branch in the color-magnitude diagram (CMD). Ages derived using this
``isochrone method'' are sensitive to uncertainties in cluster extinction
and reddening, as well as in the conversions between model (bolometric
luminosity, temperature) and observed (apparent luminosity, color)
quantities.

For NGC\,188, the values for reddening and metallicity are 
well constrained. A reddening of $E(B-V) = 0\fm087$ derived from the
\citet{schlegel98} dust maps agrees with the estimates of
\citet{shk+99} of $0\fm09 \pm 0\fm02$ from their two-color diagram.
We use $E(B-V) = 0\fm087$ throughout this paper. The metallicity of
NGC\,188 has been studied during the last two decades
\citep[e.g.][]{hobbs90,friel93,friel02,randich03,worthey03}. Results
from photometric studies agree with results from medium- to
high-resolution spectroscopy, both of which find the cluster's
metallicity to be in the range from $[Fe/H] = -0.12$ to solar. Estimates
of the cluster distance from isochrone-fitting to the CMD, have over
the past decade agreed to within a few percent, ranging from 1660\,pc
to 1710\,pc \citep{shk+99,vs04,bbs05,fts+07} with individual uncertainties
of the order of $5\%$.

NGC\,188 is a key cluster of the WIYN\footnote{The WIYN Observatory
is a joint facility of the University of Wisconsin-Madison, Indiana
University, Yale University, and the National Optical Astronomy
Observatories located at Kitt Peak National Observatoty, Arizona, USA.}
Open Cluster Study \citep[WOCS;][]{mathieu00}. As part of WOCS,
\citet{gmh+08} present a time-series radial-velocity survey of more
than 1046 stars in the field of NGC\,188 over a time baseline of
11 years. In combination with a deep proper-motion study of the
cluster \citep{pkm+03}, the WOCS spectroscopic data have securely
identified large populations of single and binary cluster members.
V\,12 was established as a double-lined spectroscopic binary and confirmed
as a cluster member early 
in the WOCS survey. The eclipsing nature of the system was discovered
by \citet{z02} in a time-series photometric survey for variable stars.
\citet{z02} recorded only one complete eclipse and classified V\,12
as an Algol-type eclipsing binary. The photometry used by \citet{z02}
located V\,12 at the foot of the giant branch in the NGC\,188 CMD.
A later photometric survey by \citet{zdz+04} placed V\,12 above the
cluster turn-off in the CMD, and estimated the period of V\,12 to be
2.8 days. Just recently, V\,12 was identified as a detached eclipsing
binary in NGC\,188 by \citet{mss+08}. They determine a period of 2.9
days.

As a spectroscopic double-lined and eclipsing binary, precise
masses and radii can be determined for the primary and secondary components
of V\,12. Because both components are near the turnoff mass for NGC\,188,
a comparison of their masses and radii to theoretical isochrones can put
tight constraints on the age of the binary, and thus its parent cluster.
Importantly, the determinations of the masses and radii are independent of
the reddening, distance, and metallicity of the system.
In addition, double-lined eclipsing binaries are excellent
primary distance indicators \citep[e.g.][]{guinan98,clausen04}.
Therefore, the analysis of eclipsing binary stars, in
particular when coupled with reliable information about the stellar
chemical composition, is important to improving our understanding of
the physical properties and evolution of stars. However, stellar evolution
models have progressed to a level where stellar parameters must be
observed with a precision of $\sim$1\% or better to provide useful
constraints \citep[e.g.][]{andersen91}. For examples of detailed
analysis of field binaries and the comparisons with stellar evolution
models see e.g. \citet{tlm+06}, \citet{ls07}, and \citet{avw08}.

While bright eclipsing binaries in relatively nearby clusters
have been observed for decades \citep{kl59,sem62,sd63,pk84,cg87,gc96},
and several studies have been published recently (see e.g. \citet{sms05}
and \citet{sc06} and references therein), it is only recently that
fainter systems in more distant clusters have been discovered and
published \citep[e.g.][]{kprt06,gch+08}.

In Section~\ref{spel} we begin the analysis of the spectroscopic
data on V\,12. We determine the orbital elements of V\,12 and give the
probabilities for it being a kinematic member of NGC\,188. Section~\ref{phot}
describes the photometric data, the light curves, and the photometric
elements of the V\,12 components. That leads to our determination of
the physical parameters for V\,12, including effective temperature,
distance, and age, in Section~\ref{absdim}. In Section~\ref{cmd} we
examine how well the isochrones that best fit the V\,12 components
represent the rest of the cluster members in the NGC\,188 CMD, and
we compare the cluster age determined via V\,12 to those determined
by the isochrone method. In Section~\ref{sc} we summarize our results
and give our conclusions.


\section{Radial velocities and spectroscopic elements of V\,12}
\label{spel}


\subsection{Spectroscopic Observations and Datareduction}
\label{specobs}

A detailed description of the spectroscopic observations and
radial-velocity measurements of stars in NGC\,188 can be found
in \citet{gmh+08}. The spectroscopic observations of V\,12 were
acquired from Sep 1996 to July 2001 with the WIYN telescope and
the Hydra multi-object spectrograph in a single 250{\AA} wide
echelle order centered on 5130 {\AA} and with a resolution of
$\sim20\,000$. For each observation three $40$ minute exposures
were combined to increase the signal-to-noise ratio of the spectrum
and to filter out cosmic ray events. The CCD bias level was determined
using the overscan-section of the chip and subtracted. A single
domeflat exposure, adjacent in time to the science exposures, was
used for the purposes of tracing and extracting the stellar spectrum,
flatfield correction, and fiber throughput correction. The three
consecutive science exposures were bracketed by Thorium-Argon (ThAr)
comparison lamp exposures used to define the pixel-to-wavelength
mapping (dispersion solution).
The DOHYDRA\footnote{http://iraf.noao.edu/tutorials/dohydra/dohydra.html}
IRAF task was used for the reduction of the WIYN/Hydra multi-spectrum
data. The signal-to-noise ratio ($S/N$) for the 16 spectra used for the
orbital solution of V12 ranges from $\sim$15-25 per resolution element
(2 pixels) of $\sim 14~km s^{-1}$. For single lined stars with slow to
moderate rotation velocities, radial velocities with a precision of
$\la 0.4~km s^{-1}$ \citep{gmh+08} were derived from the spectra via
cross-correlation with a high $S/N$ sky (solar) spectrum obtained at
evening twillight.


\subsection{Component Radial-Velocities and Orbital Elements}
\label{rvorb}

\begin{figure}
\epsscale{.80}
\plotone{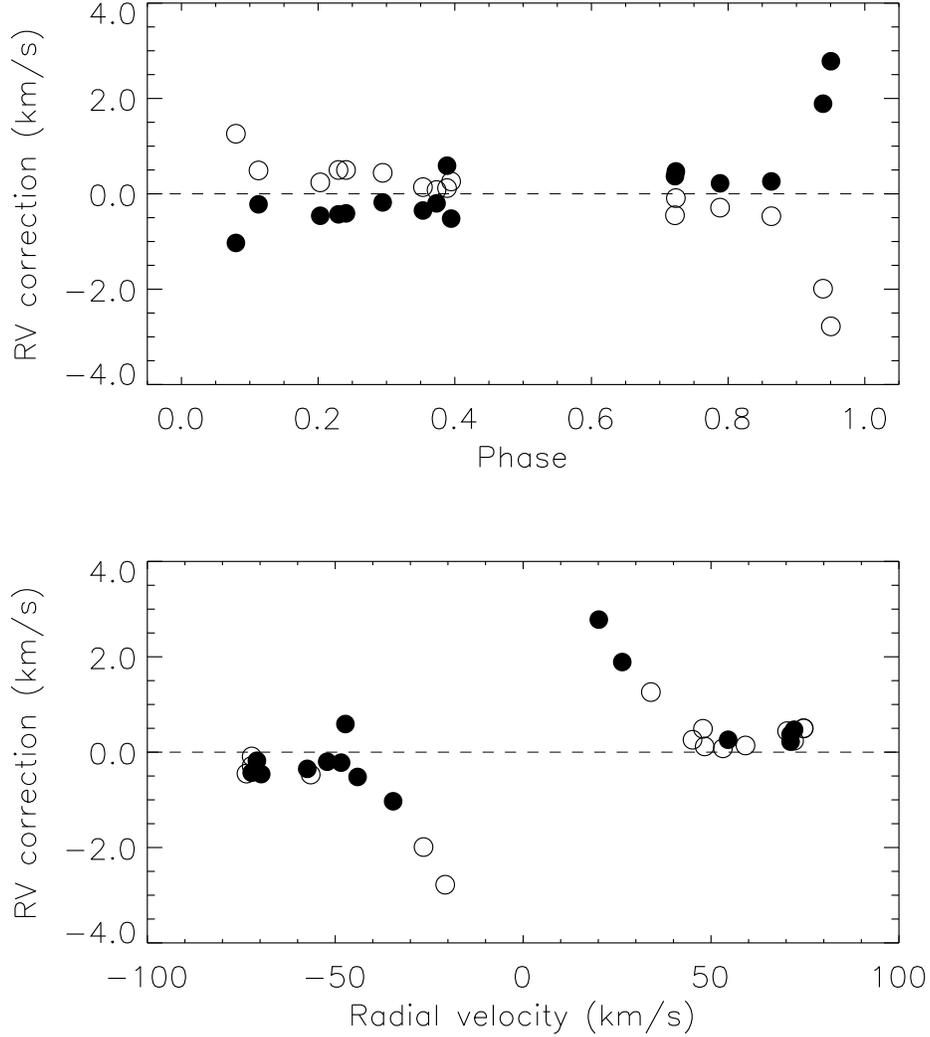}
\caption{
Systematic errors in the raw TODCOR velocities of V\,12
determined from simulations with synthetic binary spectra
(filled circles: primary; open circles: secondary).
The differences are plotted both as function of 
orbital phase (upper panel) and radial
velocity relative to the center--of--mass velocity (lower panel), and
have been applied to the measured velocities as corrections.
Phase 0.0 corresponds to central primary eclipse.
}
\label{v12_rvcor}
\end{figure}

\begin{figure}
\epsscale{1.0}
\plotone{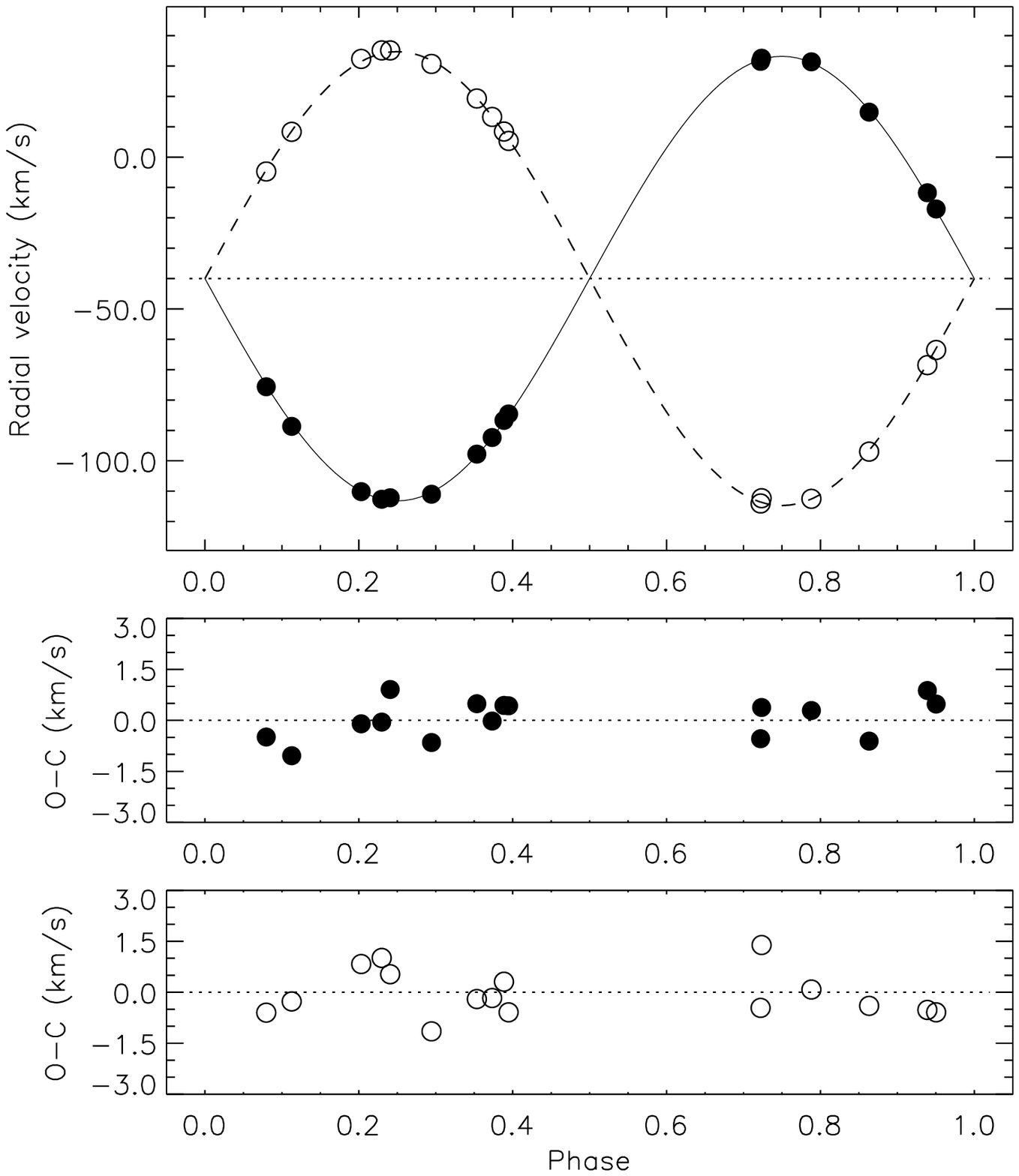}
\caption{
Spectroscopic orbital solution for V\,12 and radial velocities listed
in Table~\ref{tab:v12_rv} (filled circle: primary; open circle: secondary).
The horizontal dotted line (upper panel) represents the center--of--mass 
velocity of the system. Phase 0.0 corresponds to central primary eclipse.
}
\label{v12_sbop_cor}
\end{figure}

Because we seek maximum precision and accuracy for the V\,12 velocities,
we have used two methods to independently re-reduce the spectra and derive
velocities for both components. First, we
have determined radial velocities using the two-dimensional cross-correlation
algorithm TODCOR \citep{zucker94}. For the TODCOR analysis we have used
identical synthetic templates for the two binary components, since they are
nearly identical; see Section~\ref{absdim}. The templates were calculated from
the Synplot/Synspec\footnote{http://nova.astro.umd.edu/Synspec43/synspec.html}
tool by Hubeny \& Lanz for $T_{eff}$ = 6000 K, log($g$) = 4.0, [Fe/H] = 0.0,
$v \sin i$ = 20.0~\kms, and a resolution of R = 20\,000, adopting an
ATLAS9 model atmosphere \citep{ku93}. Preliminary spectroscopic elements
were calculated from the measured radial velocities using the method
of Lehman-Filh\'es implemented in the SBOP program \citep{e04}. The
orbital period was adopted from Equation~\ref{eq:v12_eph} (Section~\ref{tmin}).
The orbit was assumed to be circular based on an initial eccentricity of
$0.005 \pm 0.003$. Because spectral lines of the stars move in and out
of the observed spectral range with orbital phase, it is necessary to
apply corrections to the radial velocities derived from the TODCOR
algorithm (e.g. \citet{tsa+97} and \citet{tr02}). To determine such
corrections, we have constructed
synthetic binary spectra from the template, adopting for each orbital
phase the radial-velocity shifts defined by the preliminary elements.
These synthetic spectra were then analyzed by TODCOR in the same
manner as the observed V\,12 spectra. The corrections (i.e. measured $-$
calculated velocities for the synthetic binary spectra), which are shown
in Figure~\ref{v12_rvcor}, are then added to the measured V\,12 velocities
and revised spectroscopic elements are determined from SBOP. In the
case of V\,12, we find that the radial-velocity corrections are quite
significant, up to 3~\kms\ for both components. Including the corrections
improves the fit of the theoretical orbit to the radial velocities
significantly, and the effect on the absolute masses is not negligible
($\sim$2\% for both components). The final velocities including corrections
are listed in Table~\ref{tab:v12_rv}, and the corresponding spectroscopic
elements are given in Table~\ref{tab:v12_orb}. The final spectroscopic
orbital solution for V\,12 is shown in Figure~\ref{v12_sbop_cor}. Within
the errors, separate analysis of the radial velocities for
the primary and secondary components, respectively, yield identical
elements. Also, analysis performed with several other templates
($v \sin i$ = 10.0~\kms, corresponding to synchronous rotation
(see Section~\ref{specanalysis}); 200 K lower temperatures; different
spectrum code \citep{vp96} and line list \citep{kupka99}) produce
indistinguishable results.

\begin{table*}
\caption[]{\label{tab:v12_rv}
Radial velocities for V12.}
\begin{flushleft}
\begin{tabular}{llrrrr} \hline
\hline\noalign{\smallskip}
\multicolumn{1}{c}{HJD}&\multicolumn{1}{c}{Phase}      &\multicolumn{1}{c}{$RV_p$}&\multicolumn{1}{c}{$RV_s$}&\multicolumn{1}{c}{$(O-C)_p$}&\multicolumn{1}{c}{$(O-C)_s$} \\
- 2\,400\,000&                         &\multicolumn{1}{c}{$\mathrm{kms^{-1}}$}&\multicolumn{1}{c}{$\mathrm{kms^{-1}}$}&\multicolumn{1}{c}{$\mathrm{kms^{-1}}$}&\multicolumn{1}{c}{$\mathrm{kms^{-1}}$} \\
\hline\noalign{\smallskip}
50331.86241 &  0.72352 &    32.6 &  -112.3 &     0.4 &     1.4 \\
50559.93366 &  0.78823 &    31.4 &  -112.5 &     0.3 &     0.1 \\
50614.66644 &  0.20309 &  -110.2 &    32.3 &    -0.1 &     0.8 \\
50614.91145 &  0.24076 &  -112.2 &    35.1 &     0.9 &     0.5 \\
50615.77338 &  0.37328 &   -92.3 &    13.2 &    -0.0 &    -0.2 \\
50615.91243 &  0.39465 &   -84.6 &     5.3 &     0.4 &    -0.6 \\
50780.65039 &  0.72221 &    31.5 &  -114.1 &    -0.5 &    -0.5 \\
50797.99374 &  0.38866 &   -86.7 &     8.4 &     0.4 &     0.3 \\
50815.71183 &  0.11272 &   -88.7 &     8.3 &    -1.0 &    -0.3 \\
50920.96276 &  0.29447 &  -111.0 &    30.7 &    -0.7 &    -1.2 \\
50976.69696 &  0.86330 &    14.8 &   -97.0 &    -0.6 &    -0.4 \\
50996.77643 &  0.95041 &   -17.1 &   -63.5 &     0.5 &    -0.6 \\
51127.70306 &  0.07966 &   -75.6 &    -4.7 &    -0.5 &    -0.6 \\
51128.67943 &  0.22977 &  -112.7 &    35.1 &    -0.1 &     1.0 \\
51175.01367 &  0.35341 &   -97.8 &    19.3 &     0.5 &    -0.2 \\
52121.94531 &  0.93896 &   -11.7 &   -68.5 &     0.9 &    -0.5 \\
\noalign{\smallskip}
\hline
\end{tabular}
\end{flushleft}
\end{table*}

\begin{table}   
\caption[]{\label{tab:v12_orb}
Spectroscopic orbital solution for V12.
$T$ is the time of central primary eclipse.}
\begin{flushleft}    
\begin{tabular}{lr} \hline   
\hline\noalign{\smallskip}    
Parameter            & \multicolumn{1}{c}{Value} \\ 
\noalign{\smallskip}
\hline
\noalign{\smallskip}    
Adjusted quantities:            &   \\ 
$T$~(HJD$-$2\,400\,000)          &$ 50906.0389  \pm 0.0028 $\\
$K_p$~(\kms)                     &$ 73.2 \pm 0.2 $  \\
$K_s$~(\kms)                     &$ 74.7 \pm 0.2 $  \\
$\gamma$~(\kms)                  &$-40.0 \pm 0.1 $  \\
\noalign{\smallskip}  
Adopted quantities:             &     \\
$P$~(days)                      &  6.5042969  \\
$e$                             &  0          \\ 
\noalign{\smallskip}  
Derived quantities:             &      \\
$M_p \sin^3i~\mathrm{(M_{\sun})}$       & $ 1.102  \pm 0.006  $ \\
$M_s \sin^3i~\mathrm{(M_{\sun}})$       & $ 1.080  \pm 0.005  $ \\
$q (\frac{M_s}{M_p})$                   & $ 0.980 \pm 0.008 $ \\
$a_p \sin i$~($10^6$~km)                & $ 6.547 \pm 0.019 $ \\
$a_s \sin i$~($10^6$~km)                & $ 6.685 \pm 0.019 $ \\
$a \sin i~\mathrm{(R_{\sun})}$          & $19.011  \pm 0.040  $ \\
\noalign{\smallskip}  
Other quantities               &      \\
$N_{obs}$                      &    16 \\
Time span (days)               &  1790 \\
$\sigma_p$~(\kms)              &  0.65 \\
$\sigma_s$~(\kms)              &  0.77 \\
\noalign{\smallskip}  
\hline
\end{tabular}            
\end{flushleft}            
\end{table}

Secondly, we have applied spectral disentangling. This technique solves
simultaneously for the contributions of the components to the observed
composite spectra, and for the Doppler shifts in the component spectra.
We have applied the method introduced by \citet{ss94} using a version
of the original code revised for the Linux operating system. It assumes
a constant light level outside eclipses, which is fulfilled for V\,12.
Dedicated IDL\footnote{http://www.ittvis.com/ProductServices/IDL.aspx}
programs were applied to remove cosmics ray events and other defects,
and for normalization of the spectra before these analysis.
Spectral disentangling is applicable because sufficient (16) WIYN spectra
(S/N $\approx$ 15-25) are available giving good phase/velocity coverage
outside eclipses. We have assumed the ephemeris given in
Equation~\ref{eq:v12_eph} (Section~\ref{tmin}) and fitted for the velocity
semi-amplitudes of the spectroscopic orbit for each component. We obtain
$K_p$ = 73.0~\kms\ and $K_s$ = 74.7~\kms, i.e. excellent agreement with
the results given in Table~\ref{tab:v12_orb}. We have not quoted
uncertainties for the semi-amplitudes since it is not clear how to
calculate robust values in the disentangling analysis \citep{hynes98}.
The disentangled spectra have S/N ratios of $\sim$60 and cover 200\AA.


\subsection{Cluster Membership Probabilities for V\,12}
\label{mem}

The center of mass radial velocity of V\,12 is $-40.0 \pm 0.1\kms$
(Table~\ref{tab:v12_orb}). From the WOCS radial-velocity survey of
NGC\,188, \citep{gmh+08} report an average cluster velocity of $-42.36
\pm 0.04~\kms$. The center of mass velocity of V\,12 is 2.5$\sigma$
from the fitted cluster mean, corresponding to a radial-velocity
membership probability of 50\%. A 97\% astrometric membership
probability was determined for V\,12 from the proper motion study
of \citet{pkm+03}.


\section{Photometry of V\,12}
\label{phot}

The photometric data for V\,12 consist of $V$ and $I_{\rm C}$
CCD observations obtained at the 2.56m Nordic Optical Telescope (NOT) 
July 2004 - April 2005, at the 1.2m Flemish Mercator Telescope 
April 2005, and at the 0.6m telescope at Bialkow
observatory in Poland February 2005 - April 2005. At the NOT,
either StanCam or ALFOSC was used; we refer the reader to the telescope
homepages for further information.
Table~\ref{tab:phot_data} gives a summary
of the data obtained. 

\begin{table}
\caption[]{\label{tab:phot_data}
Summary of photometric data obtained for V12.
}
\begin{flushleft}
\begin{tabular}{lrrrr} \hline
\hline
Telescope           & $N_V$ & $N_I$ & $t_{exp}(V)$ & $t_{exp}(I)$ \\
\hline
NOT/StanCam         &  131  & 118   & 120          &  90 \\
NOT/ALFOSC          &  197  & 189   &  60          &  45 \\
Mercator            &    0  &  40   &   -          & 180 \\
Bialkow             &   51  &  36   & 300          & 200 \\
\hline
\end{tabular}
\end{flushleft}
\end{table}

At each telescope the flat fields used in the
data reduction were obtained during evening twilight.
All photometry was carried out with DAOPHOT/ALLSTAR \citep{stetson87} and
DAOGROW \citep{stetson90} and transformed to a common coordinate system
using MATCH and MASTER (P. Stetson, private comm.).  For each frame we
produced a point-spread function (PSF) using the brightest stars in the
field and subsequently carried out aperture photometry in large apertures
(neighbor stars subtracted) using the NEDA routine provided with DAOPHOT.
Subsequently the aperture photometry was fed into DAOGROW to obtain the
final large-aperture magnitudes. We found that this procedure gave slightly
better photometric precision than using profile-fitting photometry.


\subsection{Transformation to $V$ and $I_{\rm C}$}
\label{trans}

For the analysis of the light curve it is important that all
data in a given band are on the same photometric system.
For this purpose we used the
photometry provided by \citet{smv04} which gives calibrated
$UBVRI$ photometry for a large fraction of the stars in NGC\,188. 
We note that $R$ and $I$ are on the Cousins system.
The stars in common with \citet{smv04} were then used as
local standards and the following transformation equations was adopted
(for both $V$ and $I$):

\begin{equation}
\label{eq:phot_trans}
m_{\rm obs} = M_{\rm std} + \alpha\,\times\,(B-I)_{\rm std} + \beta 
\end{equation}

Capital letters denote calibrated magnitudes and lower case refers to
instrumental magnitudes. For each combination of telescope, instrument,
and filter, the values of $\alpha$ and $\beta$ were found by the following
procedure: 1) We fitted eq.~\ref{eq:phot_trans} to the data from each
frame (excluding V\,12 which is variable); 2) We calculated the average
value of $\alpha$, as it is expected to be constant in time; 3) Using
the average value for $\alpha$ the values of $\beta$ were re-determined
as the final frame zeropoint ($\beta$).

Using these values we calculated $V$ and $I_{\rm C}$ magnitudes for V\,12.
In the calculation of the zeropoints stars with $(B-I)$ in the interval
between 1.2 and 2.6 were used. Most of the stars had a color within
$\pm$0.3mag of V12. Thus the zeropoints
are essentially based on stars of very similar color to V\,12, ie. stars
similar to the turnoff color of NGC\,188.

Before combining the photometry and adjusting zeropoints
(see Section~\ref{lcfin}), we formed mean values of the out-of-eclipse
observations of V\,12 from ALFOSC and StanCam in order to obtain a higher
precision than obtained by \citet{smv04}. This is important for
the determination of effective temperatures and distances. RMS errors of
about 0.005 magnitudes were obtained, compared to 0.013 ($V$) and 0.018 ($I$)
for \citet{smv04}. The mean $V$ magnitudes from the two instruments
agreed, whereas a difference of about 0.02 magnitudes was seen in
$I$. We adopt the following out-of-eclipse results for V\,12: $V = 14.745
\pm 0.005$, $I = 13.951 \pm 0.010$, $V-I = 0.794 \pm 0.011$.


\subsection{V\,12 light curves}
\label{lcfin}

When combining calibrated photometry from several detector/filter
combinations the photometric scale and zeropoints are always a concern.
We therefore compared the phased light curves to check their quality.
We therefore calculated the phased light curve to check its quality.
The $V$ photometry showed good consistency
between the various sources with typical offsets from the average of
$\pm0\fm01$. However the situation for the $I$ filter was somewhat
poorer, with ALFOSC and StanCam data showing a consistent offset of
0\fm025mag. We have not been able to track down 
a definitive cause for these inconsistencies
but note that \citet{sbg03} carried out an
extensive analysis of a large dataset for the old open cluster NGC 6791
and found that combining calibrated photometry from different
telescope/filter/CCD combinations and different nights can exhibit
scatter of the order $\pm0\fm02$.
We note that the $I$ flat field of ALFOSC contains relatively large
gradients near the center of the field where V\,12 is located in our
images. This instrument is a combined focal-reducer and
spectrograph and the gradient seen is probably due to sky concentration.
We speculate that this is related to the offsets between the ALFOSC and
StanCam data.
Offsets of similar size were found for the Mercator and Bialkow data in I.

Since the internal precision of our data is significantly better than
0\fm01 we decided to adjust the zeropoints on a nightly basis for both
the V and I data in order to produce a phased light curve with the
smallest amount of scatter. To do this we first decided to adopt the
values of V and I from \citet{smv04} as our out-of-eclipse
magnitudes for V\,12 since this is well
calibrated and multi-epoch photometry (at random phases for V\,12). For
each night with photometric data out (or partly out) of eclipse we
adjusted the zeropoint to the values of \citet{smv04}. Using
the phased data, small offsets were then applied to the remaining
light curve segments.

It is important to emphasize that the depths of the eclipses 
of the final light curves are the same
as observed with the different instruments. The data from Bialkow,
NOT/StanCam and NOT/ALFOSC cover parts of the eclipse which include
the central eclipse and out-of-eclipse data, and the observed eclipse
depths in these agree to 0\fm01 to 0\fm02. We also note that for the
analysis of the light curve the absolute level of the out-of-eclipse
part does not have any influence. This only enters into the determination
of the absolute magnitude of the stars (or equivalently the cluster
distance modulus). The out-of-eclipse data from NOT show a scatter
around a constant level of 3.4 mmag for both the V and I filter.

\begin{figure}
\epsscale{1.00}
\plotone{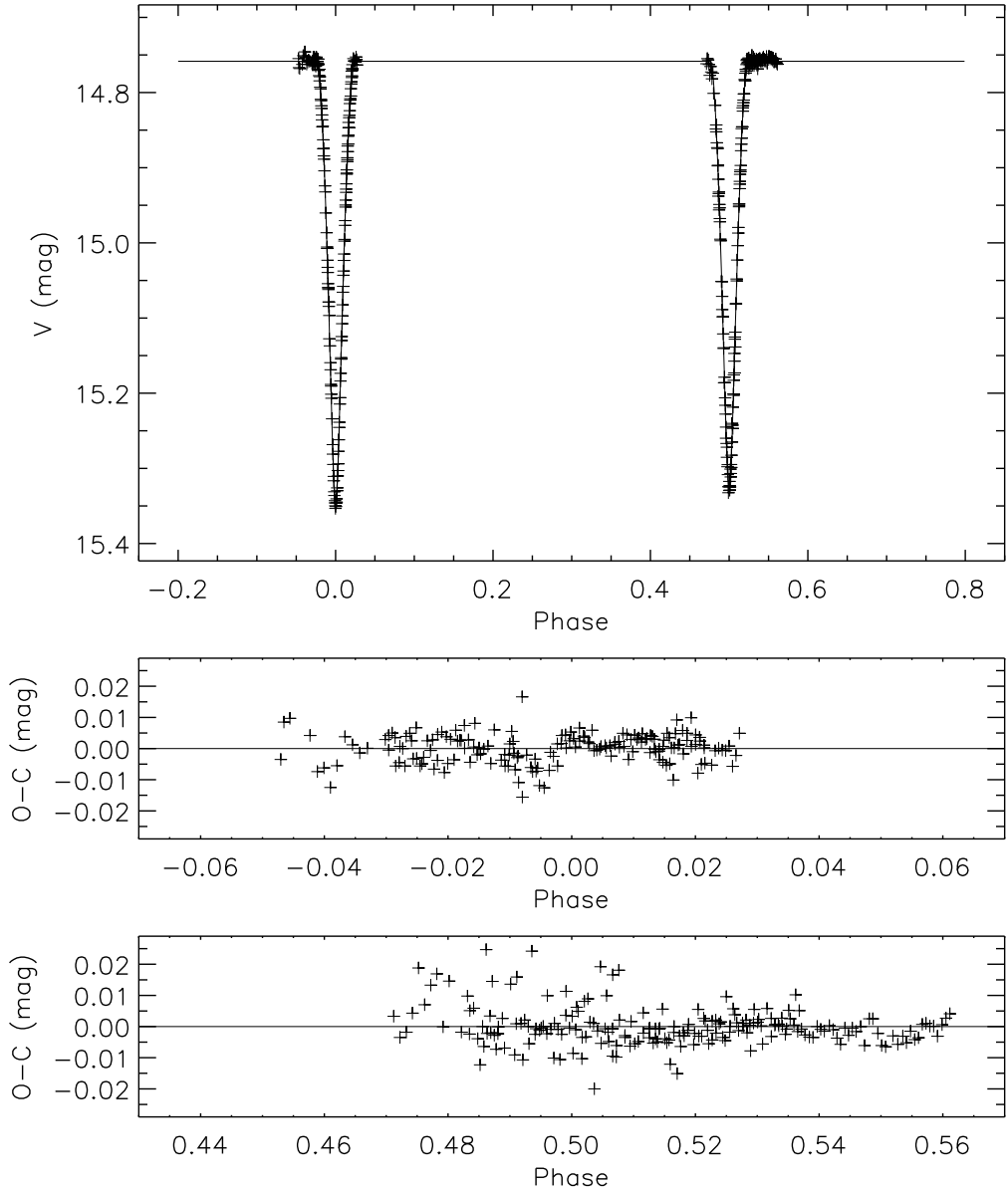}
\caption{Observed V light curve of V\,12 (upper panel, crosses) and 
the theoretical EBOP light curve (line) corresponding to the photometric 
solution presented in Table~\ref{tab:v12_ebop} assuming linear limb darkening
coefficients by \citet{vh93}.
(O-C) residuals of the observations from the theoretical light curve
are shown in the two lower panels.}
\label{lc_V}
\end{figure}

\begin{figure}
\epsscale{1.00}
\plotone{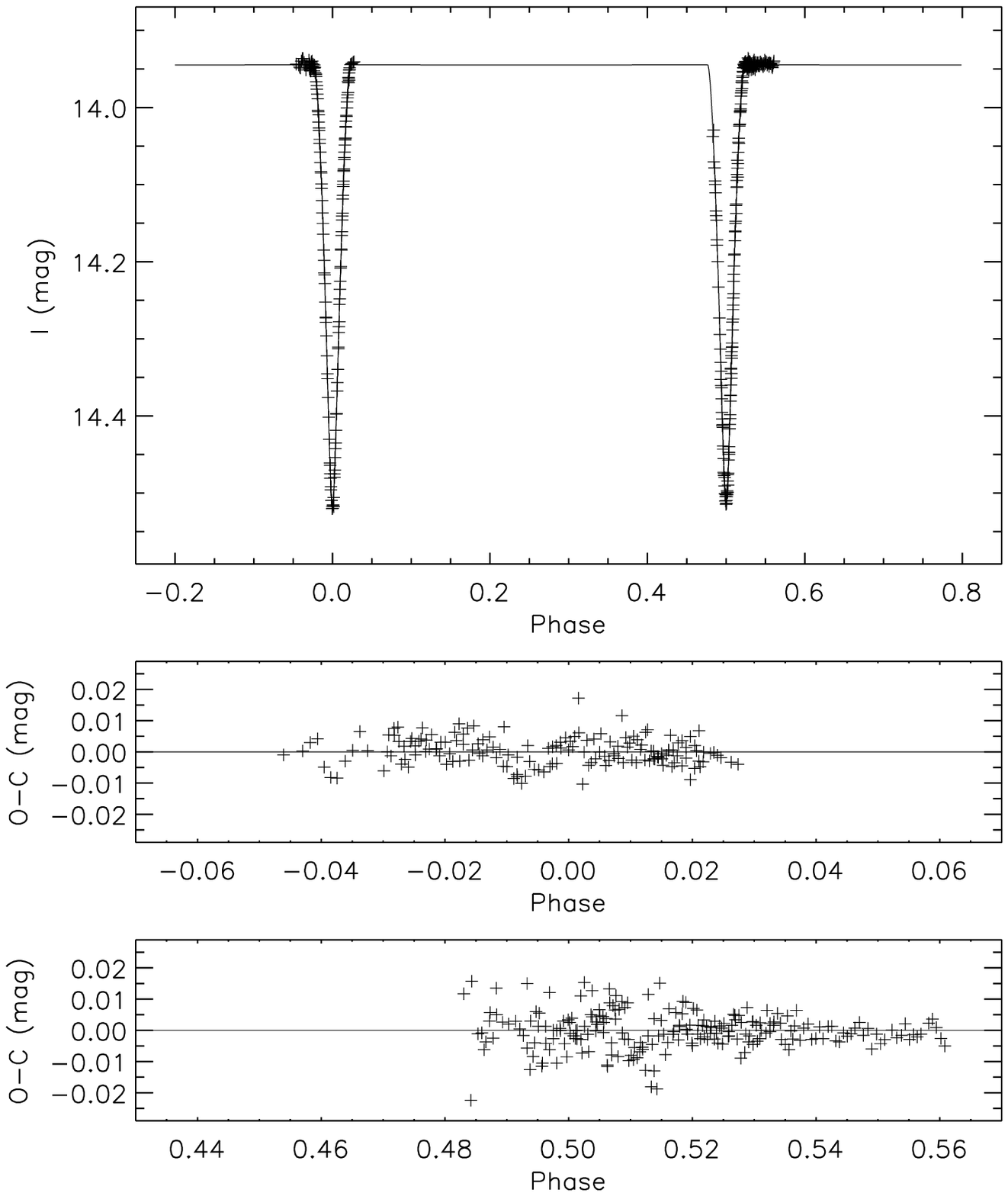}
\caption{Observed I light curve of V\,12 (upper panel, crosses) and the
theoretical EBOP light curve (line) corresponding to the photometric solution
presented in Table~\ref{tab:v12_ebop} assuming linear limb darkening
coefficients by \citet{vh93}.
(O-C) residuals of the observations from the theoretical light curve
are shown in the two lower panels.}
\label{lc_I}
\end{figure}

The light curves are shown in Figs. \ref{lc_V} and \ref{lc_I}.
A total of 379 and 383 measurements were obtained in $V$ and $I$, 
respectively \footnote{Light curve tables are available electronically}. 
Both eclipses have been covered on several nights, and there are enough
data points outside eclipses to define the maximum light level because
V\,12 is so well detached that
essentially no out-of-eclipse variations are present.
The eclipses are narrow and of almost identical depths
(about 0.6 mag in V). They are of the same duration, and secondary
eclipse occurs at phase 0.5, consistent with a circular orbit.

As a final note we point out that the V\,12 field is uncrowded and the
stars are bright. This means that it should be a fairly easy task to
obtain even better data than presented here. To avoid the transformation
difficulties encountered here this should be done using a single
telescope/detector/filter combination for each band.

\subsection{Times of minimum and ephemeris}
\label{tmin}

\begin{table*}
\caption[]{\label{tab:v12_tmin}
Times of minima for V12. 
O-C values for primary (P) and phases for secondary (S) minima 
are calculated for the ephemeris given in Eq.~\ref{eq:v12_eph}.}
\begin{flushleft}
\begin{tabular}{llcrc} \hline
\hline
HJD-2400000   & r.m.s.    & Type &   O-C/Phase&  Band       \\    
\hline
50906.03886   &0.00276    & -    &  0.00003   & Spectroscopic orbit \\
53221.56770   &0.00052    & P    & -0.00083   & I           \\ 
53221.56854   &0.00053    & P    &  0.00001   & V           \\ 
53229.61958   &0.00022    & P    & -0.00051   & I           \\
53229.62044   &0.00011    & P    &  0.00035   & V           \\
53312.62875   &0.00019    & P    &  0.00007   & I           \\
53312.62954   &0.00015    & P    &  0.00086   & V           \\
51605.25      &0.01       & S    &  0.49989   & Zhang et al (2001) \\
53452.47085   &0.00028    & S    &  0.49997   & I           \\
53452.47109   &0.00010    & S    &  0.50000   & V           \\
53465.48005   &0.00047    & S    &  0.50006   & I           \\
53465.48001   &0.00060    & S    &  0.50005   & V           \\
\hline
\end{tabular}
\end{flushleft}
\end{table*}

From the V and I light-curve observations, six times of primary
minimum and four times of secondary minimum have been derived. They
are given in Table~\ref{tab:v12_tmin} together with the time of primary
minimum obtained from the spectroscopic orbit (see Section~\ref{spel})
and the approximate time of secondary minimum by \citet{zdz+04}. 
When possible, the method of \citet{kv56} was applied, otherwise a second 
order polynomial was fitted to the observations.

An unweighted  linear least squares fit to the times of primary minimum 
yields the following ephemeris:

\begin{equation}
\label{eq:v12_eph}
\begin{tabular}{r r c r r}
{\rm Min \, I} =  & HJD 2453299.62009 & + & $6\fd 5042969$ &$\times\; E$ \\
                  &      $\pm  25$&   &       $\pm 18$ &             \\
\end{tabular}
\end{equation}

\noindent
which we adopt for the light-curve and radial velocity analysis.
Within errors, the times of 
secondary minimum give the same period, and applying weights leads
to nearly identical results. We find no evidence for eccentricity of
the orbit of V\,12.


\subsection{Photometric elements}
\label{pe}

\begin{table*}
\caption[]{\label{tab:v12_ebop}
Photometric solutions for V12 from the EBOP code. A photometric scale
factor (the magnitude at quadrature) and the phase of primary eclipse
were included as free parameters. Linear limb darkening were kept at
theoretical values or left free. COO = Claret (2000), VH93 = Van Hamme
(1993). The errors quoted for the adjusted parameters are the $formal$
errors determined from the iterative least squares solution procedure.
}
\begin{flushleft}
\begin{tabular}{lrrrrrr} \hline
\hline\noalign{\smallskip}
Band                   &  $V$       &   $V$    &     $V$    &   $I$    &      $I$  &    $I$   \\ 
Limb darkening         &  C00       & VH93     & Free       &  C00     &  VH93     & Free \\
\hline\noalign{\smallskip}            
$i$ \, (\degr)       &  88.55     &  88.62   &  88.65     &  88.57   & 88.62     & 88.64\vspace{-0.8mm}\\
                     & $\pm 1$    & $\pm 1$  &  $\pm 4$   & $\pm 1$  & $\pm 1$   & $\pm 3$ \\

$r_p$                &  0.0756    & 0.0755   &  0.0755    &  0.0742  & 0.0742    & 0.0743\vspace{-0.8mm}\\
                     &  $\pm 7$   & $\pm 7$  &  $\pm 8$   &  $\pm 8$ &$\pm 10$   & $\pm 11$ \\

$r_s$                &  0.0724    & 0.0719   &  0.0716    &  0.0731  & 0.0726    & 0.0723 \\

$k = r_s/r_p$        &  0.958     & 0.952    &  0.949     &  0.985   & 0.978     & 0.973\vspace{-0.8mm}\\
                     &  $\pm17$   & $\pm18$  & $\pm20$    &  $\pm23$ & $\pm26$   & $\pm29$ \\

$r_p + r_s$          &  0.1480    & 0.1474   &  0.1471    &  0.1473  & 0.1469    & 0.1466 \\

$u_p = u_s$          &  0.67      & 0.59     &  0.55      &  0.52    & 0.43      & 0.39\vspace{-0.8mm}\\
                     &            &          & $\pm4$     &          &           & $\pm 4$ \\

$y_p = y_s$          &  0.38      & 0.38     &  0.38      &  0.30    & 0.30      &  0.30          \\

$J_s/J_p$            &  0.9823    & 0.9824   &  0.9824    &   0.9954 & 0.9957    & 0.9960\vspace{-0.8mm}\\
                     &  $\pm28$   & $\pm28$  & $\pm 28$   &  $\pm 26$& $\pm 26$  & $\pm 26$             \\

$L_s/L_p$            &  0.901     & 0.890    &  0.884     &   0.967  & 0.953     &  0.943          \\

$\sigma$ \, (mag.)   &  0.0058    & 0.0057   &  0.0057    &   0.0055 & 0.0054    &  0.0054     \\
\noalign{\smallskip}            
\hline
\end{tabular}            
\end{flushleft}            
\end{table*}

\begin{table}            
\caption[]{\label{tab:v12_phel}
Adopted photometric elements for V12.
The individual flux and luminosity ratios are based
on the mean stellar and orbital parameters.
}
\begin{flushleft}             
\begin{tabular}{lrr}             
\noalign{\smallskip}             
\hline             
\noalign{\smallskip}                                
$i$              & $88{\fdg}62 \pm 0{\fdg}05$ &\\
$r_p$            & $0.0749 \pm 0.0010$ &\\ 
$r_s$            & $0.0722 \pm 0.0010$ &\\ 
\noalign{\smallskip}             
\noalign{\smallskip}             
                 & $V$    & $I$ \\
\noalign{\smallskip}             
$J_s/J_p$        & 0.981  & 0.996 \\   
                 & $\pm15$&$\pm14$ \\
$L_s/L_p$        & 0.913  & 0.927  \\  
                 & $\pm36$&$\pm50$\\ 
\noalign{\smallskip}             
\hline             
\end{tabular}             
\end{flushleft}            
\end{table}

\begin{figure}
\epsscale{1.00}
\plotone{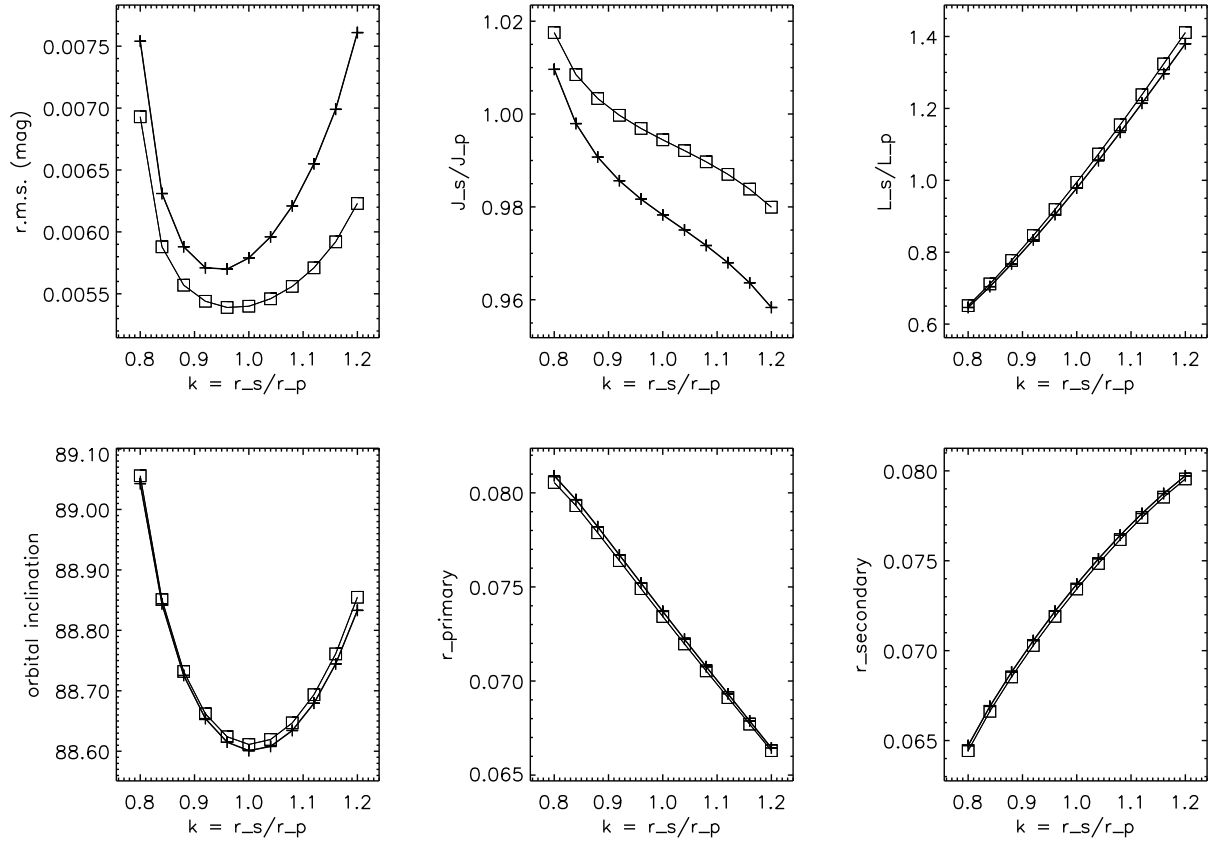}
\caption{
EBOP solutions for V\,12 for a range of fixed $k$ values. Linear
limb-darkening coefficients by \citet{vh93} were adopted. The upper left
panel shows RMS errors of the fit to the observations. Symbols are:
cross $V$; square $I$.
}
\label{v12_ebop_vh}
\end{figure}

\begin{figure}
\epsscale{1.00}
\plotone{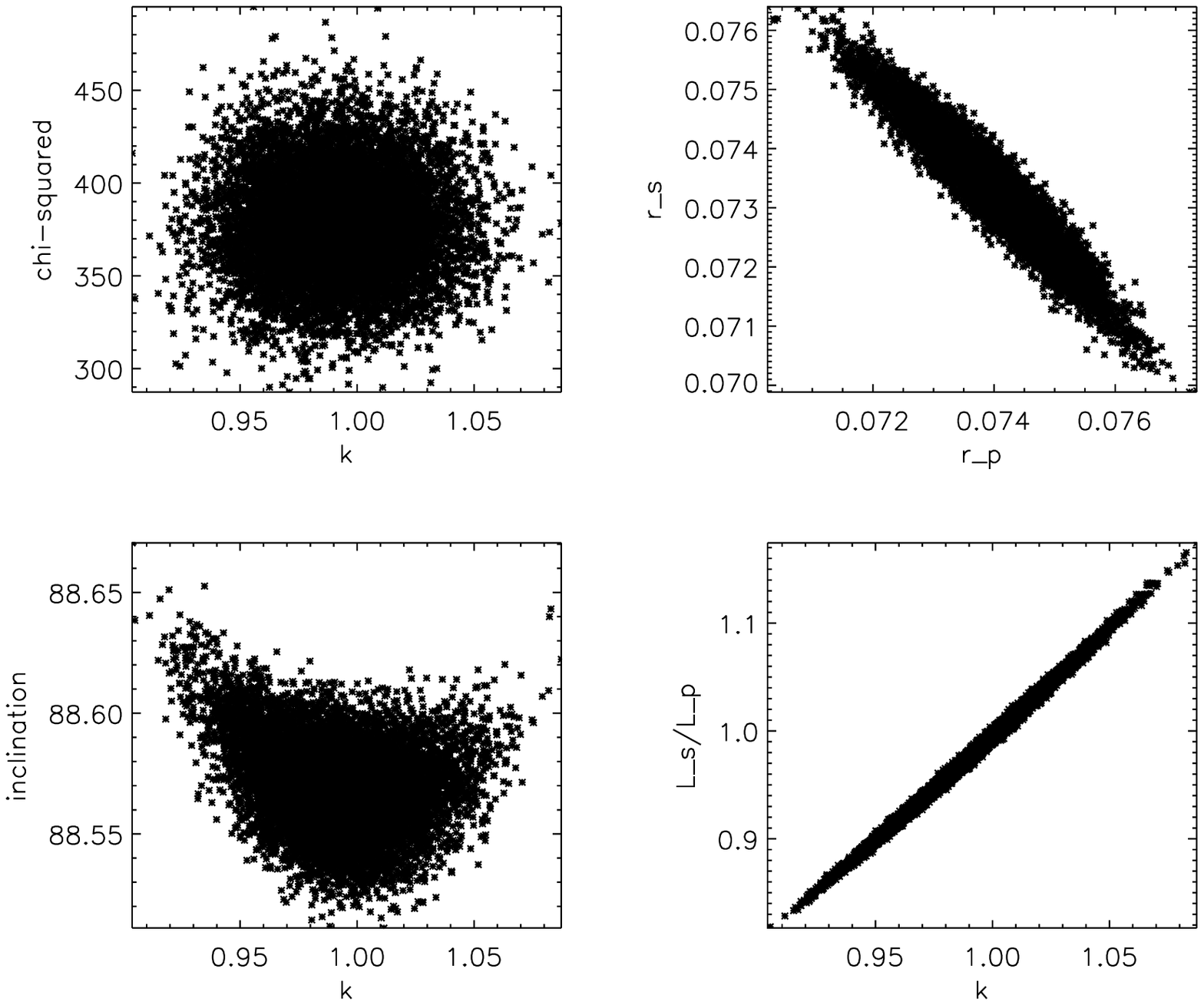}
\caption{
Best fitting parameter values for the 10\,000 synthetic V\,12 $I$ light curves
created for the Monte Carlo analysis. Linear limb darkening coefficients
by \citet{vh93} were adopted.
}
\label{v12_mc_i}
\end{figure}

The choice of binary model for light-curve analysis depends strongly
on the relative radii of the components and the irradiation between
them. In our case, the system is well detached with relative radii
below 0.1, the components are little deformed (oblateness of 0.0006)
and reflection effects are small. Therefore, the simple Nelson-Davis-Etzel
model \citep{nd72,etzel81,martynov73}, which represents the deformed stars 
as biaxial ellipsoids and applies a simple bolometric reflection model, 
is expected to be adequate.

We have adopted the corresponding EBOP code \citep{pe81}, supplemented
by an extended version JKTEBOP including Monte Carlo simulations
for the analysis and the assignment of realistic errors for the
photometric elements; see e.g. \citet{sms04,szm+04}.

We have analyzed the $V$ and $I$ light curves independently, and
for each band identical weights have been assigned to all observations.
A photometric scale factor (the magnitude at quadrature) was always included
as an adjustable parameter, and the phase of primary eclipse was allowed to
shift from 0.0. The mass ratio between the components was kept at the
spectroscopic value. Gravity darkening coefficients corresponding to
convective atmospheres were applied in accordance with the effective
temperatures derived below, and the simple bolometric reflection
model build into EBOP was used. Linear limb darkening coefficients were
either assigned from theoretical calculations \citep{vh93,c00} according
to the effective temperatures, surface gravities, and abundances, or
included as adjustable parameters.

In tables and text on photometric solutions we use the following symbols:
$i$ orbital inclination;
$r$ relative radius;
$k =  r_s/r_p$;
$u$ linear limb darkening coefficient;
$y$ gravity darkening coefficient;
$J$ central surface brightness;
$L$ luminosity.

EBOP solutions for V\,12 are shown in Table~\ref{tab:v12_ebop}.
The following parameters were adjusted: $i$, $r_p$, $k$, and $J_s/J_p$,
and as seen their formal errors are quite small.
For each band, the solutions for the different assumptions on limb darkening
agree well, and also differences between the $V$ and $I$ solutions are
small. The adjusted linear limb darkening coefficients are close to those
by \citet{vh93}, whereas those by \citet{c00} are about 0.1 higher in
both bands.

In order to assign realistic errors to the photometric elements,
we have supplemented the analysis with EBOP solutions for a range
of $k = r_s/r_p$ values near 1.00, which are compared in
Fig.~\ref{v12_ebop_vh}, and 10\,000 JKTEBOP Monte Carlo simulations
in each band; see Fig.~\ref{v12_mc_i} for details. The adopted photometric
elements listed in Table~\ref{tab:v12_phel} are the weighted mean
values of the EBOP solutions adopting the linear limb darkening
coefficients by \citet{vh93}. The final radiative parameters were
derived from a solution that uses the weigthed average of the geometric
properties. Errors are based on the Monte Carlo
simulations and comparison between the $V$ and $I$ solutions. 
We find that at phase 0.0, about 80\% of the $V$ light from the
primary component is eclipsed, and at phase 0.5 about 85\% of the
$V$ light of the secondary is blocked out.


\section{Physical parameters for V\,12}
\label{absdim}

\begin{table}   
\caption[]{\label{tab:v12_absdim}
Astrophysical data for V12. We have assumed $T_{eff\sun} = 5780$ K,
$B.C._{\sun} = -0.08$, and $M_{bol\sun} = 4.74$. We have adopted 
$E(B-V) = 0.087 \pm 0.010$, $E(V-I) = 1.31 \times E(B-V)$,
and $A_V = 3.1 \times E(B-V)$. 
}
\begin{flushleft}    
\begin{tabular}{lrr} \hline    
\noalign{\smallskip}    
\hline    
\noalign{\smallskip}    
                     &    Primary       &    Secondary      \\ 
\noalign{\smallskip}    
\hline    
\noalign{\smallskip}    
Absolute dimensions:          &                   &                 
 \\ 
$M/M_{\sun}$                  &$1.103 \pm 0.007$  &$1.081 \pm 0.007$
\\ 
$R/R_{\sun}$                  &$1.424 \pm 0.019$  &$1.373 \pm 0.019$ 
\\ 
$\log g$ (cgs)                & $4.174 \pm 0.012$ & $4.196 \pm 0.013$
\\ 
 & & \\ 
Photometric data:             &                   &                 
\\
$V$       &  $15.449 \pm 0.021$  &  $15.548 \pm 0.023$ \\     
$V_0$     &  $15.180 \pm 0.037$  &  $15.278 \pm 0.039$ \\     
$I$       &  $14.663 \pm 0.030$  &  $14.746 \pm 0.032$ \\
$(V-I)$   &  $ 0.786 \pm 0.014$  &  $ 0.803 \pm 0.014$ \\
$(V-I)_0$ &  $ 0.672 \pm 0.019$  &  $ 0.689 \pm 0.019$ \\
          &                 &               \\
 
 
$T_{\mbox{\scriptsize eff}}\,$      &  $5900  \pm 100$ &   $5875  \pm 100$ \\ 
$M_{\mbox{\scriptsize bol}}\,$      &  $3.88  \pm 0.08$&   $3.98  \pm 0.08$ \\
$\log L/L_{\sun}$ & $0.34 \pm 0.03$ &    $0.30 \pm 0.03$ \\
$B.C.$            & $-0.06$         &    $-0.06$ \\
$M_V$ &             $ 3.94 \pm 0.08$&   $ 4.04 \pm 0.08$ \\
                 &                  &             \\
$(m-M)_V$        &$11.507 \pm 0.085$& $11.504 \pm 0.086$ \\   
$V_0-M_V$        &$11.237 \pm 0.090$& $11.235 \pm 0.091$ \\   
Distance \, (pc) &$1768   \pm 74   $& $1766   \pm 74   $ \\
\noalign{\smallskip}            
\hline
\end{tabular}            
\end{flushleft}            
\end{table}                                  

\begin{table*}
\caption[]{\label{tab:v12_teff}
Effective temperatures for the combined light of V12 derived from 
recent calibrations.  
AAM = \citet{aam96},
C   = \citet{castelli99},
VC  = \citet{vc03},
RM = \citet{rm05},
MJR = \citet{mjr06}, and
CLF = \citet{clf06}.
Extinction ratios from Table 1 in \citet{rm05} were adopted. 
De-reddened colour indices are given below the observed values.


}
\begin{flushleft}
\begin{tabular}{rrrrrrrr} \hline
\hline
$(B-V)$&$(V-R)$&$(V-I)$&$(R-I)$&$(V-J)$&$(V-H)$&$(V-K_s)$ & Calibration\\
\hline
0.687  &0.427  &0.795  &0.367  &1.291  &1.660  &1.658 &\vspace{-0.8mm}\\
$\pm16$&$\pm18$&$\pm11$&$\pm20$&$\pm29$&$\pm33$&$\pm31$& \\
0.600  &0.375  &0.682  &0.306  &1.103  &1.442  &1.423 &\vspace{-0.8mm}\\
\hline
 5837  &       &       &       &       &       &      & AAM \\
 5980  & 5705  & 5899  & 6146  &       &       &      & C\\
 5887  & 5650  & 5890  &       &       &       &      & VC\\
 5846  & 5616  & 5780  & 5986  & 5871  & 5690  & 5883 & RM\\
       &       &       &       &       &       & 5927 & MJR \\
 5958  & 5659  & 5791  & 6011  & 5874  & 5804  & 5969 & CLF\\
\hline
\end{tabular}
\end{flushleft}
\end{table*}

Absolute dimensions for the components of V\,12 were calculated from the
spectroscopic and photometric elements given in Tables~\ref{tab:v12_orb}
and \ref{tab:v12_phel}. As seen in Table~\ref{tab:v12_absdim}, masses
and radii for the only slightly different components have been determined
to an accuracy of about 0.6\% and 1.4\%, respectively. Individual $V$ and
$I$ magnitudes are included, as calculated from the combined magnitudes
outside eclipses (Section~\ref{trans}) and the luminosity ratios
(Table~\ref{tab:v12_phel}). For interstellar reddening and absorption 
corrections, we have used $E(B-V) = 0\fm087 \pm 0\fm010$, $E(V-I) = 1.31
\times E(B-V) = 0\fm114$, and $A_V = 3.1 \times E(B-V) = 0\fm27$.
For $I_C$ (8140\AA), the relations by \citet{cardelli89} give the 
same $E(V-I)/E(B-V)$ ratio, and for the color of V\,12, the $E(V-I)/E(B-V)$ 
relation by \citet{dean78} leads to the same $(V-I)$ reddening.


\subsection{The systemic and component effective temperatures}
\label{teffs}

Effective temperatures $T_{eff}$ were derived from the recent
calibrations by \citet{aam96}, \citet{castelli99}, \citet{vc03},
\citet{rm05}, \citet{mjr06}, and \citet{clf06}. Because the surface
fluxes of the components are nearly identical, the two stars have
almost the same $T_{eff}$; the empirical visual flux calibration by
\citet{dmp80} with bolometric corrections by \citet{flower96} yield
a difference of only 25~K. We have therefore applied all available
color indices for the combined light from Section~\ref{trans}, \citet{smv04},
and from 2MASS\footnote{{\scriptsize\tt http://www.ipac.caltech.edu/2mass}}
$JHK_s$ photometry \citep{cutri03}, and the individual $(V-I)_0$ indices
listed in Table~\ref{tab:v12_absdim}. Results from the combined indices
are presented in Table~\ref{tab:v12_teff}. For each color index, the
temperatures derived from the different calibrations agree quite well.
Differences in the derived temperatures between indices is, in part,
due to uncertainties of the indices, which cause changes of 50-100\,K,
except for $\sim$250\,K in $(R-I)$. The temperatures based on $(V-R)$
and $(R-I)$ would increase by about 120~K and decrease by about 200~K,
respectively, if the possible zero point error in $R$ of $\fm0.02$
(noted by \citet{smv04}) is applied. A somewhat better agreement between
the temperatures is then obtained from the different indices.

For [Fe/H] = 0.00, the individual $(V-I)_0$ indices listed in
Table~\ref{tab:v12_absdim} give temperatures of 5818\,K and 5750\,K
\citep{rm05}, 5928\,K and 5858\,K \citep{vc03}, and 5829\,K and 5765\,K
\citep{clf06} for the primary and secondary components, respectively.
Formally we obtain temperature differences between the components higher
than 25~K, as derived from the $V$ surface flux ratio, but the uncertainties
of the indices, the reddening, the metal abundance, and the calibrations
themselves add up (in quadrature) to individual temperature uncertainties
of about $\pm 100~K$. Using [Fe/H] = -0.1 typically decreases the temperatures
by less than 100 degrees. Giving higher weight to the calibration by
\citet{vc03}, we assign 5900~K and 5875~K to the primary and
secondary components, respectively, both with uncertainties of $\pm 100~K$. 


\subsection{Analysis of disentangled spectra}
\label{specanalysis}

Analysis of the disentangled spectra was performed to provide independent
values for the temperatures, metallicities, and rotational velocities of
the components. We cross-correlated the disentangled spectra of each binary
component against a grid of synthetic spectra\footnote{The library is based
on ATLAS9 (http://kurucz.harvard.edu/)
and the companion program SYNTHE used to compute a synthetic spectrum
from the model atmosphere and line list. The $R \sim 500,000$ synthetic
spectra were broadened to an instrumental profile equal to the WIYN
spectra.} spanning appropriate ranges in effective temperature, surface
gravity, metallicity, and rotation velocity. The best matches (highest
cross-correlation peak heights) for fixed surface gravities ($\log g$)
of 4.17 and 4.19 were obtained for temperatures of 5900\,K and 6000\,K,
metallicities of -0.14 and -0.13, and rotation velocities of $15\kms$
and $17\kms$, respectively, for the V\,12 A and B components. The same
analysis was performed for 5 twilight sky spectra from WIYN/Hydra,
resulting in a mean temperature 60\,K above solar, a mean metallicity
0.06 below solar, and a mean rotational velocity $3.5\kms$ above solar.
We estimate that the uncertainties on temperatures and metallicities
derived from WIYN/Hydra spectra from this technique are at least 150\,K
and 0.1 dex. Thus considering the systematic offsets for the Sun , the
spectral analysis of V\,12 gives results that are entirely consistent with
the results from the photometric analysis. Furthermore, the rotational
velocities of the binary components are consistent with the synchronous
rotation ($11\kms$) expected in an old circularized binary system.


\subsection{The distance and age of V\,12}
\label{dist_age}

The accurate masses and radii determined for the components of V\,12
are independent of distance, reddening, and color-temperature
transformations, and allow direct comparison with stellar models.
From the absolute dimensions of the components, we determine
identical distance moduli of $V_0-M_V = 11\fm24 \pm0\fm09$
($(m-M)_V = 11\fm51 \pm0\fm09$), corresponding to $1770 \pm 75$\,pc.
We adopt uncertainties of $\pm0.02$ for the bolometric corrections by
\citet{flower96}. From main-sequence fitting in the \citet{smv04}
$(B-V)_0,M_V$ CMD and adopting $E(B-V) = 0\fm087$, \citet{vs04}
derive $(m-M)_V$ values of $11\fm22 - 11\fm54$ for NGC\,188,
depending on the assumed chemical composition. Based on the
\citet{shk+99} $(B-V)_0,M_V$ CMD and adopting also $E(B-V) =
0\fm087$, \citet{mrrv04} get $(m-M)_V$ = $11\fm40$ from models
of solar metallicity.

To determine the age of V\,12, we have chosen to use the recent $Y^2$
models by \citet{yale04} and the latest VRSS Victoria-Regina models for
scaled solar abundances for the heavy elements \citep{vr06}. Comparisons
were done for [Fe/H] = 0.00 ((Y,Z) = (0.266,0.018)) and [Fe/H] = -0.10
((Y,Z) = (0.259,0.014)), corresponding to the range of recent values
for the cluster metallicity.
The two sets of models treat convective core overshooting differently,
but at the age of NGC\,188 overshooting is not required to match the
turnoff region where V\,12 is located. We refer the reader to the original
papers or \citet{avw08} for more details on the models and their input physics.

\begin{figure}[ht!]
\epsscale{1.0}
\plottwo{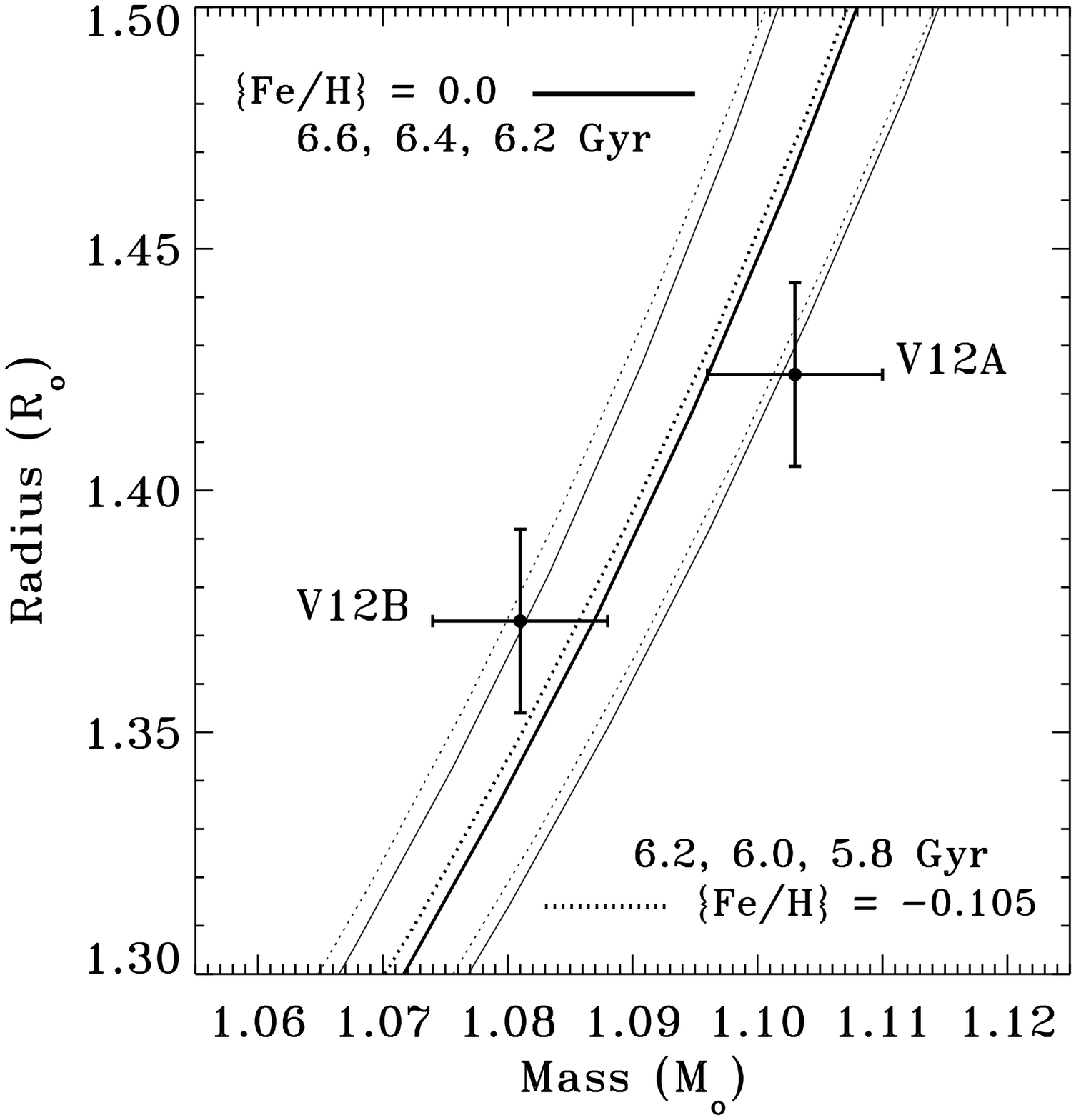}{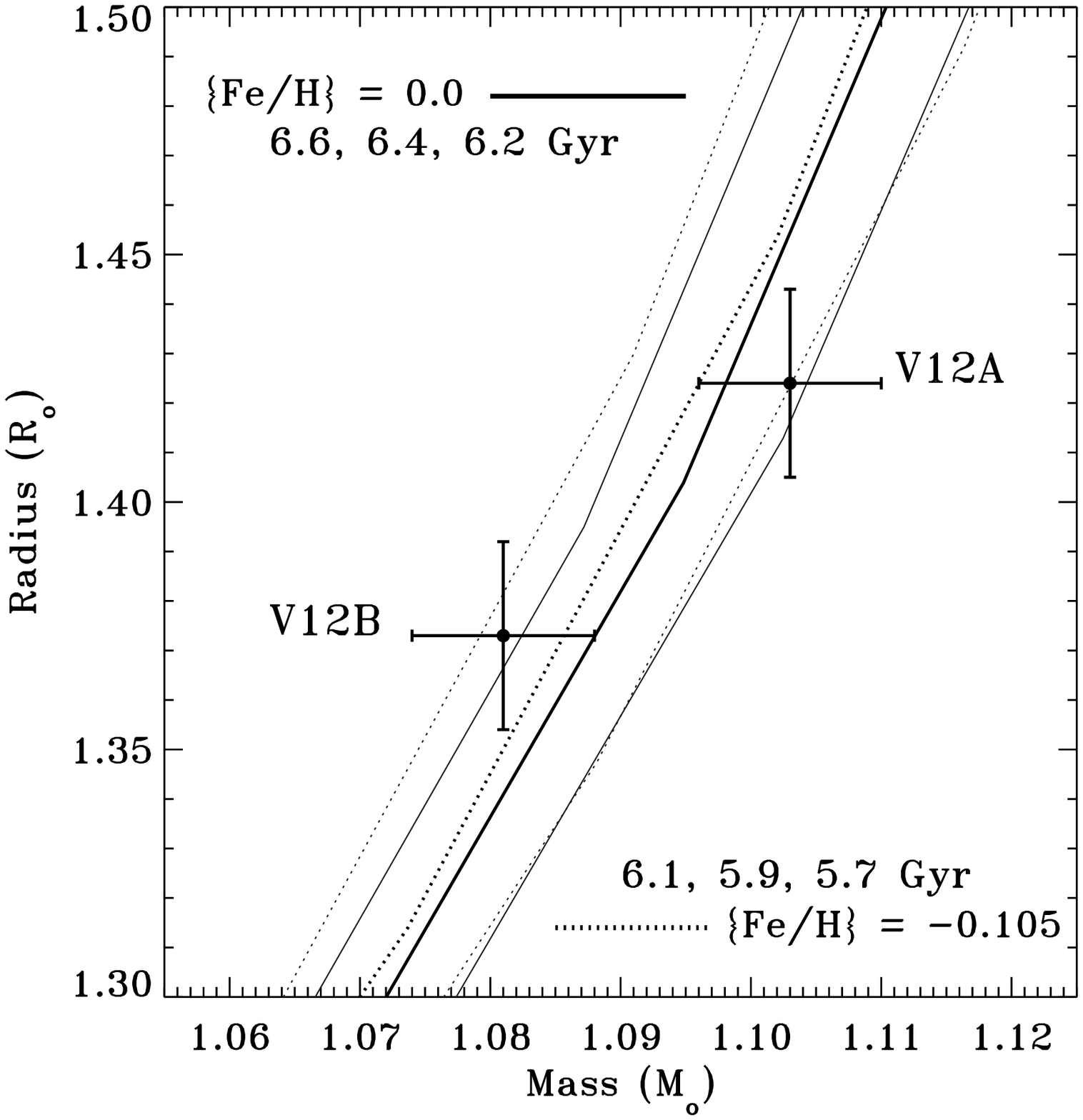}
\caption{
{\bf a)} Mass-radius diagram with V\,12 components and best fit VRSS
isochrones for [Fe/H] = 0.0 (solid line) and [Fe/H] = -0.1
(dotted line). The ages of the isochrones in Gyr are listed in
the figure in the same left-to-right order. {\bf b)} Mass-radius
diagram with V\,12 components and best fit $Y^2$ isochrones for [Fe/H]
= 0.0 (solid line) and [Fe/H] = -0.1 (dotted line). Isochrone
ages are listed as in a).}
\label{fig:v12_mr}
\end{figure}

Figure~\ref{fig:v12_mr} shows the mass-radius diagrams with the
locations of the V\,12 components and the $Y^2$ and VRSS isochrones for
relevant ages and [Fe/H] of 0.0 and -0.1. Grids of isochrones with
0.1 Gyr age separations were computed for both models using the
available interpolation routines. For [Fe/H] = 0.00, the 6.4\,Gyr
isochrones from both models, provide the best fit to the V\,12
components. For [Fe/H] = -0.1, the best-fit VRSS isochrone has
an age of 6.0 Gyr, and the best fit $Y^2$ isochrone has an age of
5.9 Gyr. Table~\ref{tab:v12_age} lists the isochrone ages for the
two different models and the two different chemical compositions considered.
We adopt an age of $6.2 \pm 0.2$\,Gyr for V\,12 and NGC\,188.
Thus, for solar metallicity, we agree with the age of 6.2 Gyr obtained
by \citet{vs04}. We also confirm the age of 6.4 Gyr
derived from solar metallicity diffusive models by \citet{mrrv04}
and the result by \citet{swp04}, which is $6.3 \pm 0.8$ Gyr for
[Fe/H] = $-0.03 \pm 0.06$. In contrast, \citet{vs04} derive a much
higher age (7.7\,Gyr) than us for [Fe/H] = -0.10. Overall, the
isochrone method applied to the NGC\,188 CMD results in a much larger
age uncertainty for V\,12 than found by our direct method over the
same abundance interval.

\begin{table}
\caption[]{\label{tab:v12_age}
Ages for V12 as determined from fitting isochrones to the binary
components in the mass-radius plane. Two different models and two
different chemical compositions are considered. Age uncertainties
due to mass and radius errors are about $\pm 0.25$ Gyr.
}
\begin{center}
\begin{tabular}{ccccc} \hline
\hline
Model         & [Fe/H]    &  $Y$   &  $Z$   & Age (Gyr) \\
\hline                            
VRSS          & 0.00    & 0.2768 & 0.0188 & 6.4 \\   
              & -0.10   & 0.2684 & 0.0150 & 6.0 \\
$Y^2$         & 0.00    & 0.2662 & 0.0181 & 6.4 \\   
              & -0.10   & 0.2589 & 0.0145 & 5.9 \\
\hline
\end{tabular}
\end{center}
\end{table}

\begin{figure}[ht!]
\epsscale{1.0}
\plottwo{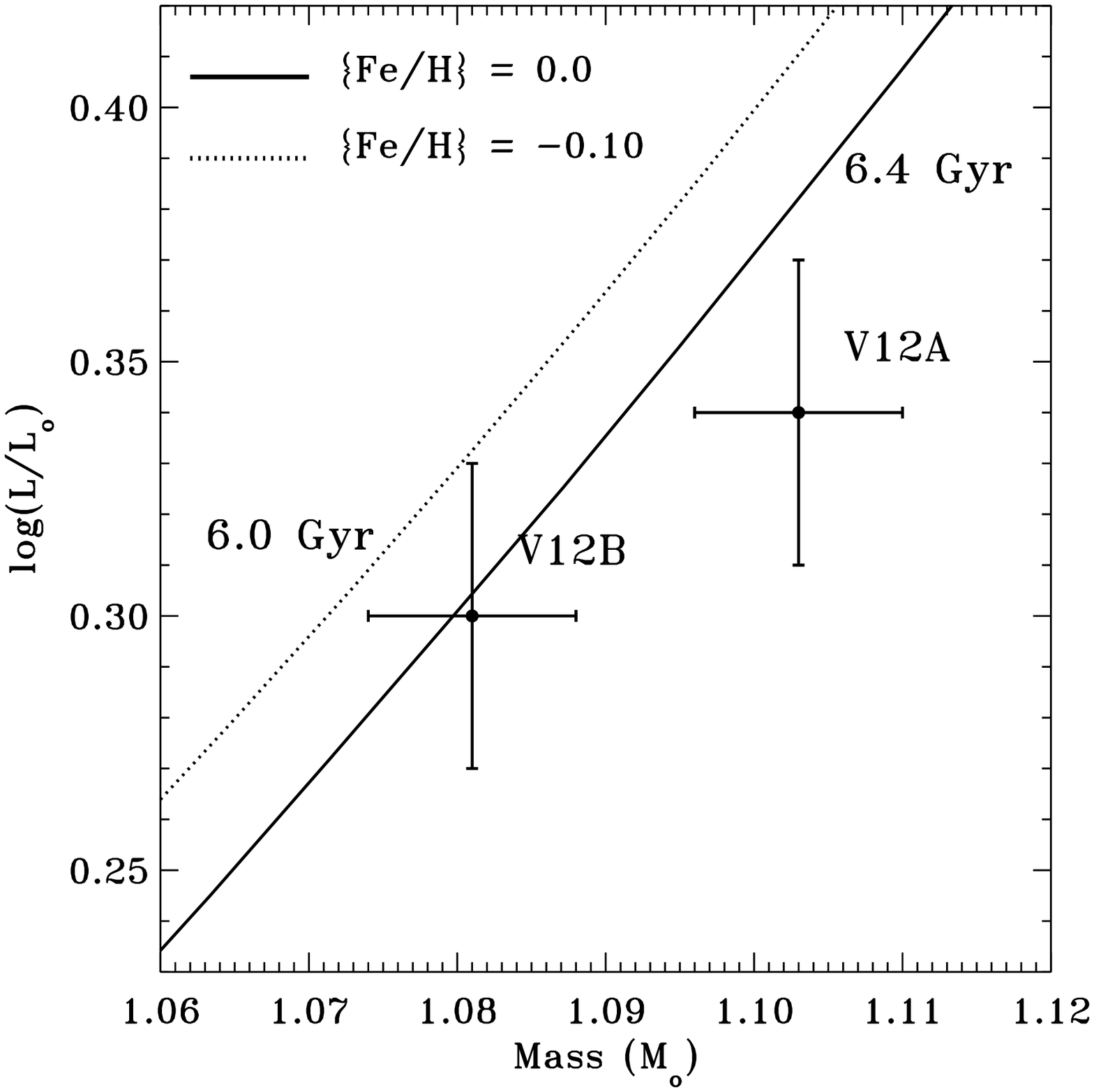}{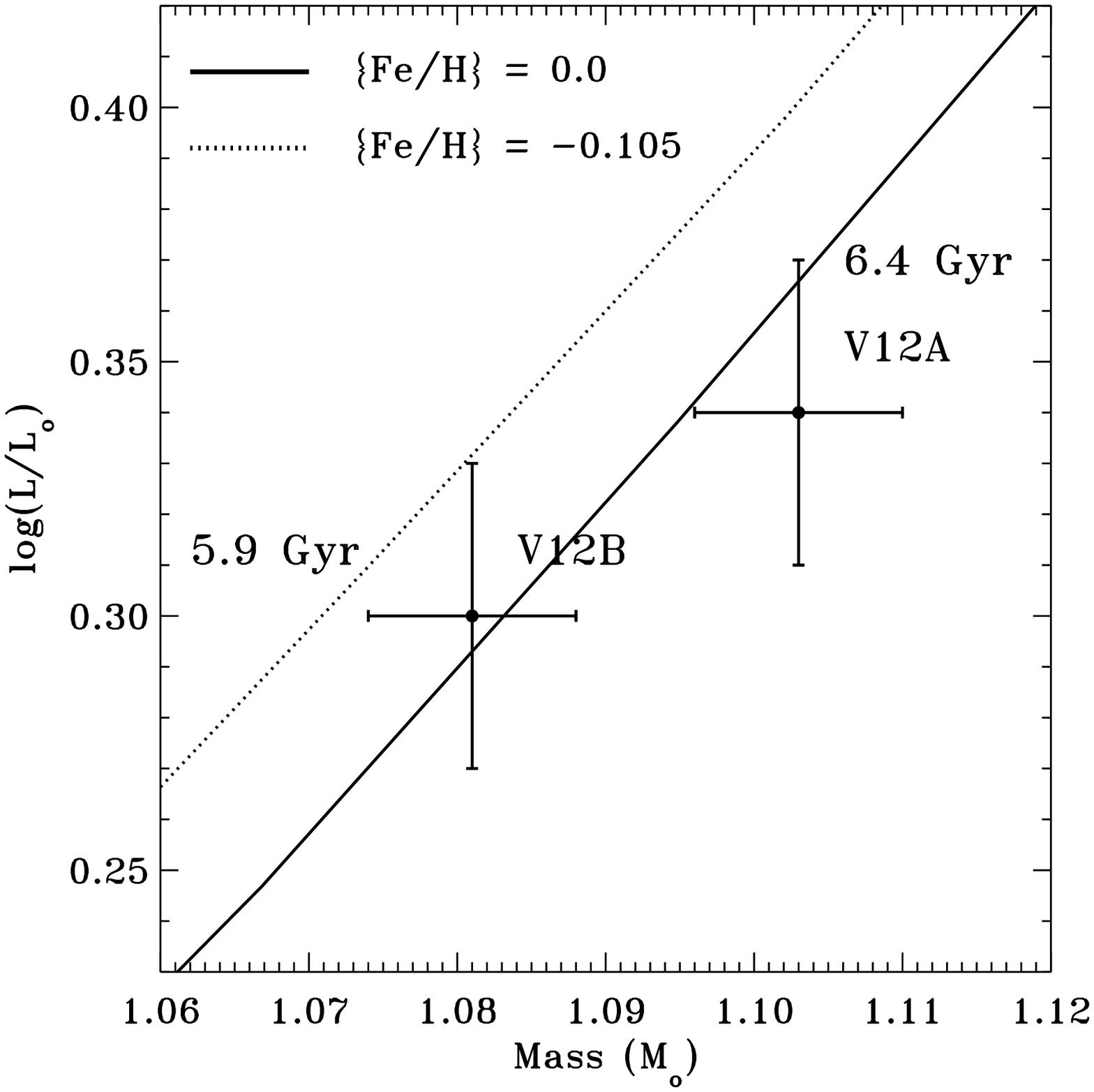}
\caption{
Mass-luminosity diagram with V\,12 components. In {\bf a)} the 6.4\,Gyr
[Fe/H] = 0.0 and 6.0\,Gyr [Fe/H] = -0.1 VRSS isochrones are shown as
solid and dotted curves, respectively. In {\bf b)} the 6.4\,Gyr [Fe/H] = 0.0
and 5.9\,Gyr [Fe/H] = -0.1 $Y^2$ isochrones are shown as solid and dotted
curves, respectively. The isochrone ages are those determined from the
mass-radius diagram.
}
\label{v12_ml}
\end{figure}

In Figure~\ref{v12_ml} we present mass-luminosity comparisons with models.
On the theoretical side, the luminosity is independent of the mixing length
assumed for convection; this is not the case for the radius. On the
observational side, the luminosity includes uncertainties for the
temperature calibration. In general, the models that best fit the
components in the mass-radius plane are too luminous, with only the
solar metallicity 6.4 Gyr $Y^2$ model marginally within the component errors.
The best fit of VRSS and $Y^2$ models to the masses and luminosities of the
binary components gives ages of 6.1\,Gyr and 6.3\,Gyr for [Fe/H] = 0.0,
and 5.3\,Gyr and 5.4\,Gyr for [Fe/H] = -0.1. To reach agreement between
the model ages determined from the mass-radius and mass-luminosity
diagrams would require an increase in the effective temperature of the
V\,12 components. The necessary temperature changes are $\sim$150\,K and
$\sim$400\,K for the [Fe/H] = 0.0 and [Fe/H] = -0.1 VRSS isochrones,
and $\sim$760\,K for the [Fe/H] = -0.1 $Y^2$ isochrone. We note that the
high sensitivity of luminosity to the effective temperature greatly
enhances the uncertainty on the luminosity, making ages derived from
the mass-luminosity diagrams less reliable.

\begin{figure}[ht!]
\epsscale{1.0}
\plottwo{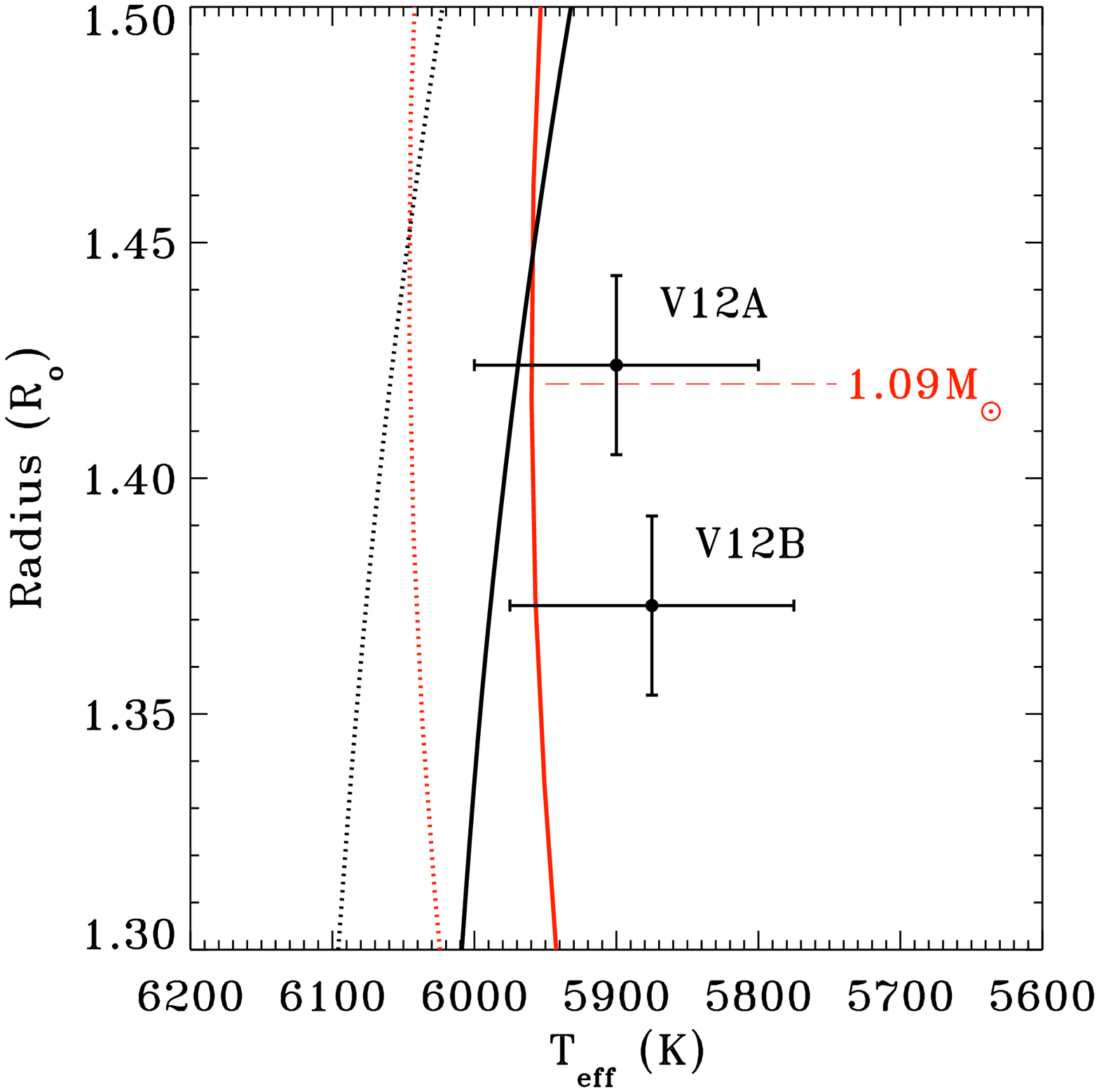}{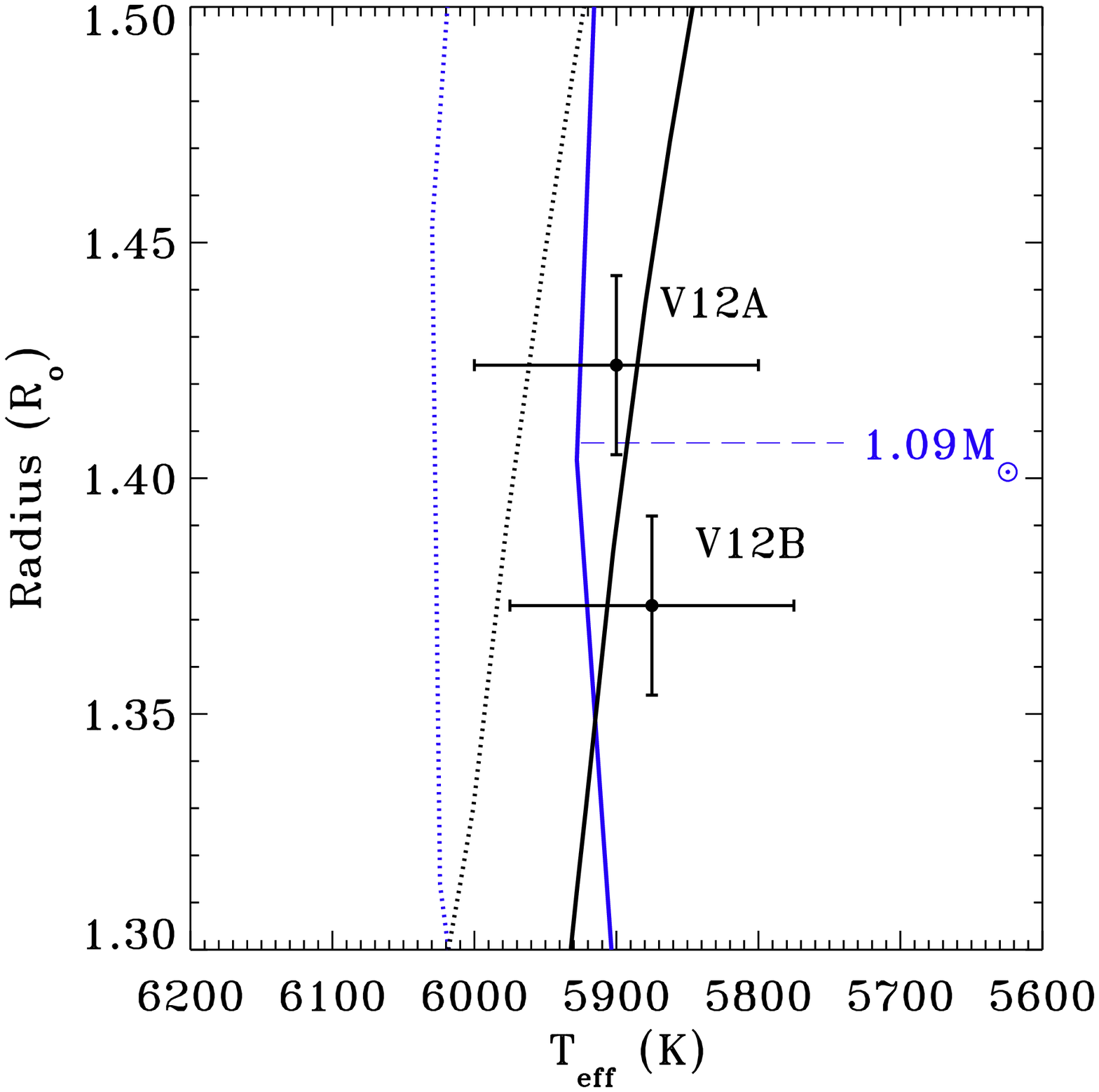}
\caption{
The effective temperature vs. radius diagram comparing the values
of the V\,12 components to those of VRSS and $Y^2$ isochrones and mass-tracks.
{\bf a)} The 6.4\,Gyr [Fe/H]=0.0 and 6.0\,Gyr [Fe/H]=-0.1 VRSS isochrones
are shown as solid and dashed curves, respectively.
The solid and dashed grey lines represent the VRSS $1.1\,M_{\odot}$
evolutionary tracks for [Fe/H]=0.0 and [Fe/H]=-0.1, respectively.
{\bf b)} Black solid and dashed curves show the 6.4\,Gyr [Fe/H]=0.0
and 5.9\,Gyr [Fe/H]=-0.1 $Y^2$ isochrones, respectively. The
$Y^2$ $1.1\,M_{\odot}$ mass-tracks are shown as in a). In both
{\bf a)} and {\bf b)} the horizontal dashed lines mark the points
on the solar-metallicity isochrones corresponding to a stellar
mass of 1.09\,$M_{\odot}$.
}
\label{v12_tr}
\end{figure}

Figure~\ref{v12_tr} presents the effective temperature vs. radius
diagram comparing the values of the V\,12 components to those of the
VRSS and $Y^2$ isochrones and $1.1\,M_{\odot}$ mass tracks. In
Figure~\ref{v12_tr}a we show the 6.4\,Gyr [Fe/H]=0.0 and the
6.0\,Gyr [Fe/H]=-0.1 VRSS isochrones as solid and dotted curves,
respectively. In Figure~\ref{v12_tr}b the 6.4\,Gyr [Fe/H]=0.0 and
the 5.9\,Gyr  [Fe/H]=-0.1 $Y^2$ isochrones are plotted as solid
and dotted curves, respectively. In both figures mass tracks
corresponding to $1.1\,M_{\odot}$ are shown as grey solid and
dashed curves for [Fe/H]=0.0 and [Fe/H]=-0.1, respectively.
For solar metallicity, the VRSS isochrone and track only marginally fit
the observed quantities of the V\,12 components, being slightly hotter,
whereas for [Fe/H]=-0.10 both are clearly too hot, especially noting
that abundance uncertainties of $\pm0.10$ are included in the temperature
errors. Similarly, the $Y^2$ isochrone and track for [Fe/H]=-0.10 are also
too hot for a proper match to the binary components, although slightly
closer in temperature than the corresponding VRSS models. The solar metallicity
$Y^2$ 6.4\,Gyr isochrone and $1.1\,M_{\odot}$ mass-track, however, provide
a very close match to the V\,12 components. We also note, that the point
on the solar metallicity 6.4\,Gyr $Y^2$ isochrone corresponding to a stellar
mass of 1.09\,$M_{\odot}$, falls appropriately between the 1.103\,$M_{\odot}$
and 1.081\,$M_{\odot}$ components.

From the mass vs. luminosity and effective temperature vs. radius
comparisons, the 6.4\,Gyr solar-metallicity $Y^2$ model provide the
best match to the observed and derived properties of the V\,12 components.
However, we discuss in Section~\ref{cmd} how a zeropoint error in the
I-band photometry from \citet{smv04}, may cause the derived temperatures
for the V\,12 components to be too cool by $\sim$100-150\,K. With a
correction for such a temperature offset, the VRSS and $Y^2$ isochrones
and mass-tracks would fit V\,12 equally well.


\section{NGC\,188 and V\,12}
\label{cmd}

\begin{figure}
\epsscale{1.00}
\plotone{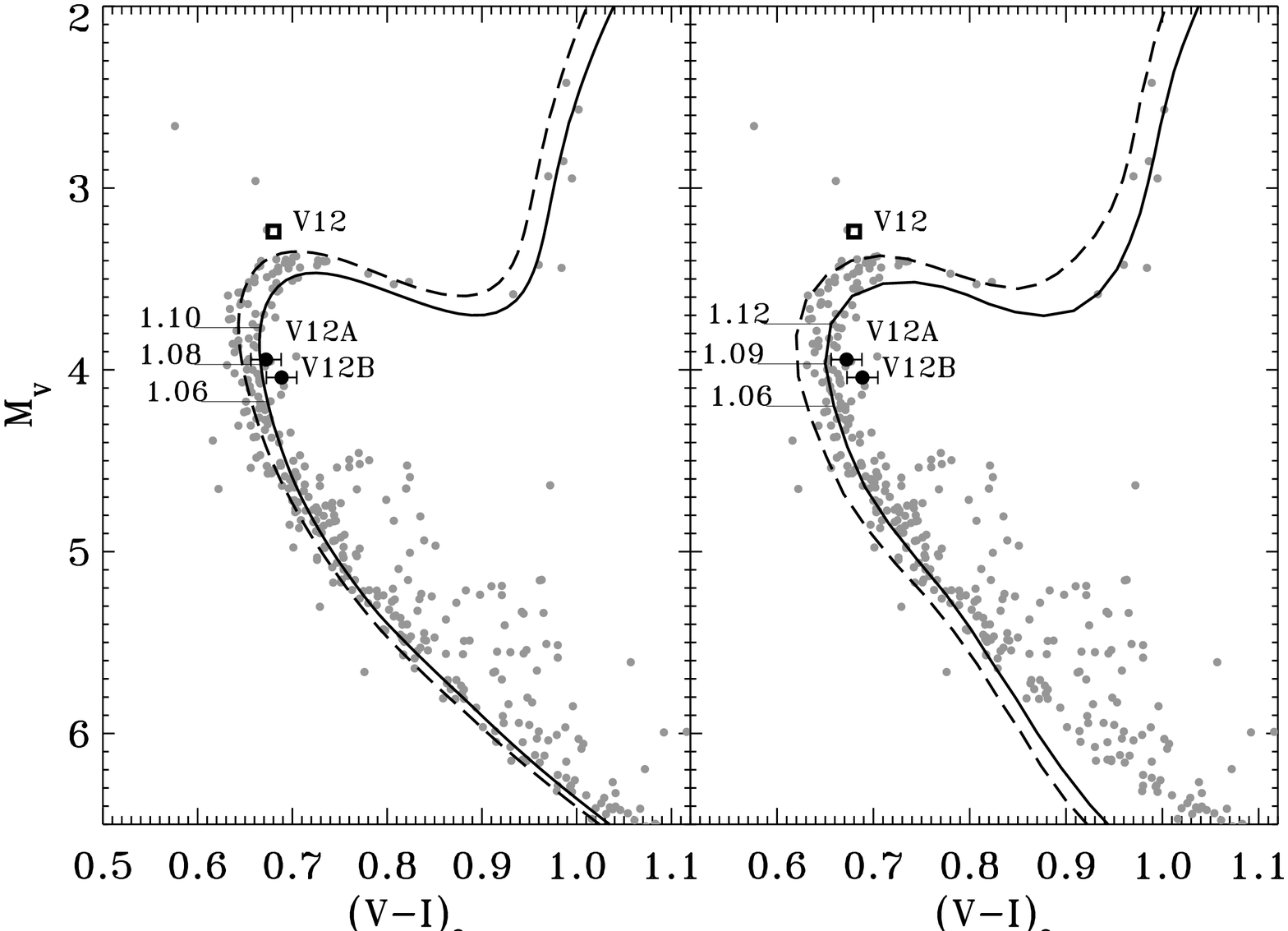}
\caption{The color-magnitude diagram for NGC\,188 based on photometry
from \citet{smv04}. All stars brighter than $M_{V}$ of 4.5 are single
radial-velocity and proper-motion members of NGC\,188 \citep{gmh+08,pkm+03}.
Stars fainter than $M_{V}$ of 4.5 are proper-motion members of NGC\,188
\citep{pkm+03}. The position of V\,12 and the individual components are
marked using the colors and magnitudes listed in Table 7.
In the left panel we show the 6.4\,Gyr [Fe/H] = 0.0 (solid)
and the 6.0\,Gyr [Fe/H]=-0.1 (dashed) VRSS isochrones. Relevant mass-points
are marked on the 6.4\,Gyr ([Fe/H] = 0.0) isochrone. In the right panel,
we display in the same manner the 6.4\,Gyr [Fe/H] = 0.0 (solid) and
the 5.9\,Gyr [Fe/H]=-0.1 (dashed) $Y^2$ isochrones.
}
\label{VIcmds}
\end{figure}

The purpose of this section is to present V\,12 in the NGC\,188 CMD,
and to investigate how well the VRSS and $Y^2$ isochrones that best
fit the V\,12 components in the mass-radius diagram, fit the rest
of the cluster's main-sequence, turn-off, and giant branch stars.
We show in Figure~\ref{VIcmds} the NGC\,188 V-I vs. V CMD using
the photometry from \citet{smv04}. All stars brighter than $M_{V}$
of 4.5 are single radial-velocity and proper-motion members of
NGC\,188 \citep{gmh+08,pkm+03}. Stars fainter than $M_{V}$ of 4.5 are
proper-motion members of NGC\,188 \citep{pkm+03}. The position of V\,12
and of its components are marked using the V magnitudes and V-I colors
listed in Table 7. Adopting the distance ($V_0-M_V = 11\fm24$) and
ages of V\,12 determined from the analysis of V\,12 in Section~\ref{dist_age},
and a reddening of 0.087, we overplot in the left panel of
Figure~\ref{VIcmds} the 6.4\,Gyr solar metallicity and the 6.0\,Gyr
[Fe/H]=-0.1 VRSS isochrones. In the right panel of Figure~\ref{VIcmds}
we have overplotted the 6.4\,Gyr solar metallicity and the 5.9\,Gyr
[Fe/H]=-0.1 $Y^2$ isochrones. For the solar metallicity isochrones we
have marked the turnoff masses nearest the V\,12 primary and secondary
components.

Both VRSS isochrones fit the main-sequence, turnoff, and giant
branch of NGC\,188 well. The $Y^2$ isochrones follow the turnoff
curvature well, but both are too blue on the lower main-sequence,
and the metal-poor $Y^2$ isochrone also falls to the blue of the giant
branch. It is clear that without the constraints on the distance
and age set by V\,12, and on the cluster metallicity by spectroscopic
studies, good fits to the cluster stars could be obtained from
a range of different cluster metallicities, distances, and ages.
For NGC\,188, this is demonstrated well by \citet{smv04}, who
show that ages between 5.9\,Gyr and 8.1\,Gyr result from the isochrone
method when plausible ranges in metallicity and distance modulus
are considered. This age-range corresponds to $\sim$33\% of their mean
cluster age of $6.8^{+1.1}_{-0.9}$\,Gyr. With V\,12 we provide a
constraint on the age of NGC\,188 with only a $\la$10\% uncertainty
due primarily to the uncertainty in the cluster metallicity. Judging
from current spectroscopic measurements and from the best fit isochrones
to the cluster, it is likely that the cluster metallicity is between
solar and = -0.1. With better determination of the cluster metallicity,
the age of V\,12 and NGC\,188 can be further improved.

We note that in the V-I colors from \citet{smv04}, V\,12 and thus
both its components are offset to the red by a few hundreds of a
magnitude relative to cluster turnoff. The offset is likely due
to a zeropoint error in the I band, because the Stetson et al.
B-V color of V12 and its components match the B-V color of the
cluster turnoff. Similar offsets for V\,12 are found in the B-V
and V-I colors in the photometry from \citet{shk+99}. Because we
calibrate our photometry using the \citet{smv04} photometry for
V\,12 (see Section~\ref{trans}), we inherit this I band
zeropoint shift. Therefore, in Figure~\ref{VIcmds}, V\,12 and both
its components display a slight red offset with respect to the cluster
turnoff. The V-I color offset correspond to $\sim$100-150\,K and
may explain, in part, the discrepancies in luminosity and effective
temperature between V12A,B and the VRSS and Y2 isochrones in
Figure~\ref{v12_ml} and Figure~\ref{v12_tr}. The color offset has
no affect on the age determination for V12 in this paper because the
age is based on the stellar masses and radii only. However,
any change in the color of V\,12 will affect the distance determined
for the binary and thus NGC\,188.


\section{Summary and conclusions}
\label{sc}

We have presented time-series photometric and spectroscopic observations
of the detached, double-lined, eclipsing spectroscopic binary V\,12
in the old open cluster NGC\,188. We have determined masses of $1.103
\pm 0.007~M_{\odot}$ and $1.081 \pm 0.007~M_{\odot}$, and radii of
$1.424 \pm 0.019~R_{\odot}$ and $1.373 \pm 0.019~R_{\odot}$ for the
binary primary and secondary components. The uncertainties
in the component masses and radii are not affected by the reddening,
metallicity, and the distance to the binary. From the absolute dimensions
of the binary components, we derive identical distance moduli of $V_0-M_V =
11\fm24 \pm0\fm09$ for both components, corresponding to $1770 \pm 75$\,pc.

By comparing the masses and radii of the binary components to the 
Victoria-Regina (VRSS) and the Yale ($Y^2$) model isochrones with [Fe/H]
= -0.1 and [Fe/H] = 0.0, we determine a mean age of 6.2\,Gyr
for V\,12 and thus NGC\,188, with an uncertainty of 0.2\,Gyr. This
uncertainty is primarily due to the 0.1 range in [Fe/H] for NGC\,188
found in the literature. Our age for V\,12 agrees with the age
determined by \citet{vs04} for a solar metallicity isochrone, and
with the the ages derived by \citet{mrrv04} and \citet{swp04}.
We find good agreement between the ages derived from both the VRSS
and $Y^2$ models. The solar-metallicity 6.4\,Gyr $Y^2$
isochrone match well the masses, radii, and effective temperatures
of both the primary and secondary components of V\,12. The corresponding
VRSS isochrone have temperatures slightly hotter ($\sim 30\,K$) at
the component radii, but still fit the components within the errors.
A systematic offset in the V-I color of V\,12 from \citet{smv04} may
explain the temperature discrepancies between the V\,12 components
and the VRSS isochrones and mass-tracks. Any future correction of
the V-I color for V\,12 will change its distance.

We use the distance and ages for V\,12, together with best estimates
for the metallicity and reddening of NGC\,188, to investigate the
locations of the corresponding VRSS and $Y^2$ isochrones relative to
cluster members in the CMD. We find that those isochrones fit the
cluster cluster giant branch, turnoff, and upper main-sequence reasonably
well. However, equally good or better fits to the cluster stars could
be obtained by adopting different model metallicities, distances, and
ages. This was demonstrated well in \citet{smv04} where plausible changes
in the model metallicity and distance result in a range of ages from
5.9\,Gyr to 8.1\,Gyr for NGC\,188 - more than 3 times that resulting
from our analysis of V\,12.

The age of V\,12 and NGC\,188 can be further improved from better
determination of the metallicity of NGC\,188 from spectroscopic
studies of main sequence, turn off, and subgiants in NGC\,188.
Additional radial-velocity measurements, and an effort to obtain
improved light curves in more bandpasses, could also provide
masses and radii with smaller errors, and thus put better constraints
on the stellar models. Finally, additional precise eclipse photometry
in e.g. the K-band can make possible an independent determination of the
distance to V12 via surface-brightness relations
\citep[e.g.][]{benedetto98,ktf+04}.


\acknowledgments
This work has been supported by NSF grant AST-0406615 to the UW-Madison,
a fellowship from the Danish Research Academy to S.M., and partial support
to S.M. from the Kepler mission via NASA Cooperative Agreement NCC2-1390.
The projects "Stellar structure and evolution - new challenges from ground
and space observations" and "Stars: Central engines of the evolution of
the Universe", carried out at Aarhus University and Copenhagen University,
are supported by the Danish National Science Research Council. F.G.
greatfully acknowledges financial support from the Danish Asterology
Science Centre (DASC) at the University of {\AA}rhus and from the
Carlsberg Foundation. We thank observers and staff at the WIYN, Nordic
Optical, Flemish Mercator, and Bialkow telescopes. We thank P. Maxted
for making his TODCOR software available and J. Southworth for many
valuable discussions and for access to is JKTEBOP code. We are grateful
to E. Sturm for providing his original disentangling code, and to him
and J.D. Pritchard for modifying it for use at Linux/Unix computer systems.
The following Internet-based resources were used in research for this
paper: the NASA Astrophysics Data System; the SIMBAD database and the
ViziR service operated by CDS, Strasbourg, France; the ar$\chi$iv
scientific paper preprint service operated by Cornell University;
This publication makes use of data products from the Two Micron All Sky
Survey, which is a joint project of the University of Massachusetts and
the Infrared Processing and Analysis Center/California Institute of
Technology, funded by the National Aeronautics and Space Administration
and the National Science Foundation.

\end{document}